\documentclass[aps,prb,twocolumn,superscriptaddress]{revtex4-2}
\usepackage{times}
\usepackage{amssymb}
\usepackage{amsmath}
\usepackage{bm}
\usepackage{float}
\usepackage{graphicx}
\usepackage{mathrsfs}
\usepackage{mathtools}
\usepackage{soul}
\usepackage{tikz}
\usepackage{verbatim}
\usepackage{xcolor}
\usepackage{epstopdf}
\usepackage[normalem]{ulem}
\usepackage[caption=false]{subfig}
\usepackage[pdftex,colorlinks,linkcolor={red!75!black},citecolor={red!75!black},urlcolor={blue!75!black}]{hyperref}

\newcommand{\mrm}[1]{\mathrm{#1}}

\newcommand{\tx}[1]{\text{#1}}

\begin{document}

\title{Edge theory of non-Hermitian skin modes in higher dimensions}

\author{Kai Zhang}
\email{phykai@umich.edu}
\affiliation{Department of Physics, University of Michigan, Ann Arbor, Michigan 48109, United States}
\author{Zhesen Yang}
\email{yangzs@xmu.edu.cn}
\affiliation{Department of Physics, Xiamen University, Xiamen 361005, Fujian Province, China}
\author{Kai Sun}
\email{sunkai@umich.edu}
\affiliation{Department of Physics, University of Michigan, Ann Arbor, Michigan 48109, United States}

\begin{abstract}
    In this paper, we establish an effective edge theory to characterize non-Hermitian edge-skin modes in higher dimensions. We begin by proposing a bulk projection criterion to straightforwardly identify the localized edges of skin modes. Through an exact mapping, we show that the edge-skin mode shares the same bulk-boundary correspondence and localization characteristics as the zero-energy edge states in a Hermitian semimetal under open-boundary conditions, bridging the gap between non-Hermitian edge-skin effect and Hermitian semimetals. Another key finding is the introduction of ``skewness,'' a term we proposed to describe the characteristic decay direction of skin mode from the localized edge into the bulk. Remarkably, we demonstrate that skewness is an intrinsic quantity of the skin mode and can be analytically determined using the corresponding cylinder-geometry bulk Hamiltonian, without requiring any boundary details. Furthermore, we reveal that in the edge-skin effect, the spectrum exhibits anomalous spectral sensitivity to weak local disturbances, a feature that crucially distinguishes it from the corner-skin effect. 
\end{abstract}

\maketitle

\section{Introduction}

Non-Hermitian band theory has been developed over the years~\cite{FuLiang2018_PRL,GongPRX2018,SatoPRX2019,Ashida2020,Kunst2021RMP,OkumaReview2023} and broadly utilized across various physical scenarios~\cite{FengLiang2017,Vincenzo2020NP,SongFei2019,FLPRL2020}. 
A central focus in this field lies on the non-Hermitian skin effect~\cite{Yao2018,Torres2018,ChingHua2019,LeeCH2019_PRL,Kai2020PRL,Okuma2020_PRL,YYFPRL2020,LiLH2020_NC,Kawabata2020,Zirnstein2021PRL,CXGuo2021PRL,XJLiu2023PRB,Frank2023PRXQ,Denner2023IOP,LeeCHReview,YFChen2022Review}, where the majority of bulk states are localized at boundaries under open-boundary conditions (OBCs). 
This phenomenon gives rise to intriguing physics unique to non-Hermitian systems, e.g., the enrichment in the band topology~\cite{LeeModel2016,Yao2018,Kunst2018_PRL,Slager2020PRL,Kai2020PRL,Okuma2020_PRL,DuanLM2021PRL}, unidirectional transport in the long-time dynamics~\cite{McDonaldPRX2018,Ashvin2019PRL,Sato2021PRL,LQPRL2022,Kai2023PRL}, and so on~\cite{Diego2019PRL,Longhi2019PRR,YYFPRL2020,Longhi2020PRL,XueWT2021PRB,LiLH2021NC,SunXQ2021PRL,Kawabata2021PRL,LuMing2021,Longhi2022PRL,XWT2022PRL,Kai2023PRB}. 
In one dimension, with the establishment of the non-Bloch framework~\cite{Yao2018,Murakami2019_PRL,ZSaGBZPRL,SFei2019PRL,DengTS2019,KawabataPRB2020,Thomale2020,Ghatak2020,XuePeng2020,XuePeng2021PRL}, the (non-Hermitian) skin modes can be accurately characterized by employing the generalized Brillouin zone (GBZ), which extends momenta from real $k$ to a set of piecewise analytic closed loops of complex variable $\beta:=e^{ik}$ on the complex plane. 

\begin{figure}[t]
    \begin{center}
    \includegraphics[width=1\linewidth]{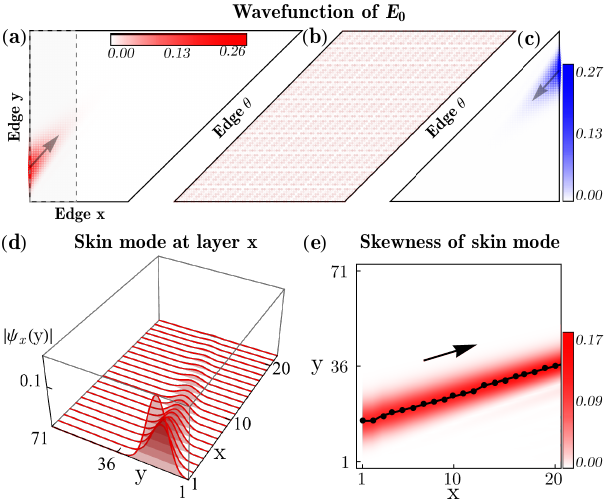}
    \par\end{center}
    \protect\caption{\label{fig:1}~Spatial distribution of wavefunction with energy $E_0{=}0.86 {-} 0.81 i$ is shown under (a)~trapezoid, (b)~parallelogram, and (c)~triangle geometries. 
    The edge-$\theta$ is oriented along $\theta = \pi/4$, and the system size in (a) is $L_y{=}71$, $L_x{=}41$. 
    The corresponding color bars represent $|\psi_{E_0}(\bm{r})|$. 
    (d)~The wavefunction components within the first 20 $x$-layers near the edge-$y$, as indicated by the gray region in (a). 
    (e)~The maximum points for each $x$-layer are selected (the black dots) and compared with the renormalized wavefunction (the red trajectory), which reveals a characteristic direction (the black arrow) of the skin mode, termed skewness. 
    The Hamiltonian parameters in Eq.(\ref{MT_Model}) are set to $t{=}2, \gamma{=}1$. } 
\end{figure}

In two dimensions, skin modes display complicated localization features due to the geometric freedom of the open boundary conditions, falling into two categories: the corner-skin and edge-skin effects. 
Phenomenologically, corner-skin modes localize at the corners of a given polygon geometry~\cite{WangZhong2018,LeeCH2019_PRL,Nori2019,LiLinhuPRL2020}, whereas edge-skin modes generally concentrate on the edges of the polygon~\cite{Kai2022NC,Wang2022NC,YQarXiv2023}. 
Both cases have been observed in recent experimental works~\cite{XDZhang2021NC,CYF2021NC,QYZhouN2023C,KunD2023PRL,WanSciB}. 
Parallel to these experimental advances, several theoretical attempts~\cite{HuiJiang2022,Murakami2022,HYWang2022,HPHu2023} have tried to extend the non-Bloch theorem into higher dimensions. 
Although the exact mathematical description and full understanding for the higher-dimensional GBZ theory are still under explored, there are many other intriguing and not yet fully understood challenges or puzzles in the higher-dimensional edge-skin effect (as illustrated in Fig.~\ref{fig:1}). 
Given this setting, finding a concise and effective edge description to understand and characterize the edge-skin modes is urgently needed. 

In this paper, we establish an effective edge theory to address the topic of edge skin effect. 
The description of edge theory includes two parts. 
First, we provide a bulk projection criterion to determine the localized edge for a generic skin mode. 
We demonstrate that edge-skin modes possess the same topological origin and spatial distribution as the exact zero-energy edge states in Hermitian semimetals with fully OBCs. 
Second, to describe how the skin mode decays from the localization edge into bulk, we define a characteristic decay direction of skin mode, termed skewness. 
It is generally believed that the localization property of skin modes depend on specific open-boundary conditions and edge details. 
Remarkably, we find that the skewness of skin mode can be precisely determined by only the non-Bloch bulk Hamiltonian with real-valued edge momenta, without requiring any boundary details. 
Here, the edge momenta specifically refer to the momenta along the direction of the localization edge. 
This finding indicates that the skewness serves as an intrinsic localization quantity of two-dimensional edge-skin modes. 
Additionally, we find that in the edge-skin effect, the corresponding cylinder-geometry spectrum is highly sensitive to weak local disturbances, such as on-site impurities, which is in sharp contrast with the  corner-skin effect. 

\section{Model with edge-skin effect}

We start with a minimal example that shows the edge-skin effect. 
The Bloch Hamiltonian is given by
\begin{equation}\label{MT_Model}
	h(\bm{k}) = 2 t \cos{(k_x+k_y)} - 2 i \gamma (1-\cos{k_x}),
\end{equation}
where ${\gamma\neq 0}$ introduces non-Hermiticity. 
We focus on a generic energy $E_0$ and consider its corresponding open-boundary wavefunction, denoted as $\psi_{E_0}(\bm{r})$, and show its spatial distribution under different open-boundary geometries, as illustrated in Figs.~\ref{fig:1}(a)-(c). 
The wavefunction exhibits the feature of geometry-dependent localization: it localizes on edge $y$ under geometries shown in Figs.~\ref{fig:1}(a)(c) but appears as an extended bulk wave under the geometry in Fig.~\ref{fig:1}(b). 

More interestingly, despite the different open-boundary geometries, the wavefunction of $E_0$ always shows the same declination of its decay direction when it moves away from the localized edge, as indicated by the black arrows in Figs.~\ref{fig:1}(a)(c). 
For clarity, we extract the wavefunction components in the first-20 $x$-layers near the edge $y$ (that is, the region inside the gray box with a dashed boundary in Fig.~\ref{fig:1}(a)) and plot them in Fig.~\ref{fig:1}(d). 
It shows that the amplitude of these layer components decreases and their centers shift as they move away from the edge layer.
To illustrate this declination behavior of the skin mode, we introduce a characteristic decay direction. 
This direction (indicated by the black arrow) is determined by the position of maximum amplitude for each layer-$x$ component of the skin mode, as shown by the black dots in Fig.~\ref{fig:1}(e). 
We term the decay direction near edge-$y$ as skewness for a given edge-skin mode. 
As a comparison, the red trajectory in Fig.~\ref{fig:1}(e) shows the wavefunction distribution near the edge $y$ after the renormalization of each layer $x$ component. 

The wavefunction localization behaviors from above numerical observations are general features and not limited to this minimal model. 
Following these numerical observations, we can pose two central questions of ``where'' and ``how'' regarding the description of edge-skin modes: 
(i) Is there a guiding principle to determine where (on which edge) skin modes localize? 
Alternatively, for a given energy $E_0$, can we straightforwardly identify the OBC geometry where the corresponding wavefunction appears as an extended bulk wave? 
(ii) How does the skin mode decay from the localized edge to bulk? Can we analytically determine the skewness for generic edge-skin modes? 
Next, we establish the edge theory including two sections to address these questions, respectively. 

\section{Edge Theory I: bulk projection criterion for edge-skin modes}
\begin{figure}[t]
    \begin{center}
    \includegraphics[width=1\linewidth]{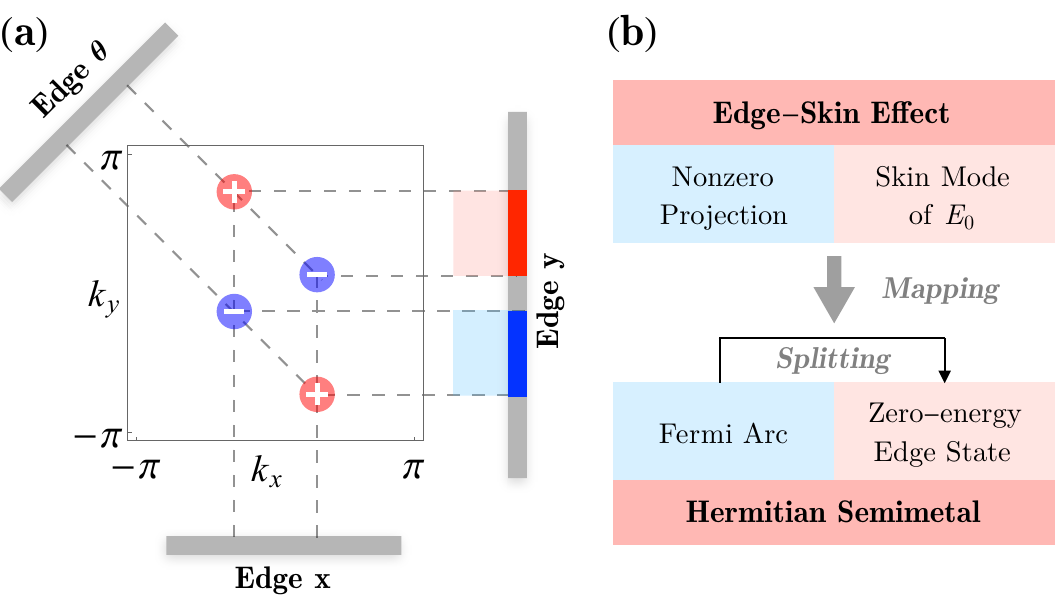}
    \par\end{center}
    \protect\caption{\label{fig:2}~(a) Fermi points of the OBC eigenvalue $E_0 = 0.86 - 0.81i$ are distributed on the Brillouin zone and their projections onto specific edges.
    (b) The mapping from the non-Hermitian reciprocal edge-skin effect to the Hermitian semimetal. } 
\end{figure}

For a generic OBC eigenvalue $E_0$, we can define the Fermi points $\bm{k}_{i}$ corresponding to this energy that satisfy $\det [h(\bm{k}_i)-E_0] = 0$. 
Note that the Hamiltonian in Eq.(\ref{MT_Model}) respects reciprocity, $h^T(\bm{k})=h(-\bm{k})$, where the superscript $T$ represents the transpose operation. 
The reciprocity of the Hamiltonian ensures the spectral property that the OBC eigenvalues are included in the Bloch spectrum (see a proof in Appendix~\ref{AppendixA}). 
Therefore, the equation $\det [h(\bm{k})-E_0]=0$ always has solutions of real $\bm{k}$. 
Each Fermi point can be assigned a topological charge based on the spectral winding number along an infinitesimal path of $\bm{k}$ that surrounds this Fermi point (see details in Appendix~\ref{AppendixB}). 
For a given edge, say edge-$y$, we can define the corresponding projective spectral winding number along the normal direction of the edge as:
\begin{equation}\label{MT_ProWinding}
    w(E_0,k_{y}) =\frac{1}{2\pi} \int_{-\pi}^{\pi} d k_{x}\,\partial_{k_{x}}\mathrm{arg}\,\mathrm{det}[h(k_{x},k_{y})-E_0 ].
\end{equation}
The projective winding number $w(E_0,k_{y})\neq 0$ when $k_y$ is in the range of the projection of these Fermi points, as indicated by the colored region in Fig.~\ref{fig:2}(a); and $w(E_0,k_{y})=0$ when $k_y$ is beyond this region. 
We specify the edge with a nonzero projective spectral winding number as a nonzero-projection edge. Thus, edge-$y$ is a nonzero-projection edge. 
As shown in Fig.~\ref{fig:1}(a), the four Fermi points of $E_0$ have no projection onto edge $x$ and edge $\theta$. 
Therefore, for all $k_x$ and $k_{\theta}$, both the projective winding numbers $w(E_0,k_{x})$ and $w(E_0,k_{\theta})$ are zero. 
These edges are thereby termed zero-projection edges. 
For a generic OBC eigenvalue $E_0$, the corresponding zero-projection and nonzero-projection edges can be readily identified by the projection of the Fermi points on these edges. 

Now, we state the bulk projection criterion for the localization of the wavefunction with energy $E_0$: 
The wavefunction appears as an extended bulk wave when the open-boundary geometry is composed of zero-projection edges [Fig.~\ref{fig:1}(b)]; while it localizes at the nonzero-projection edge when the open-boundary geometry includes one nonzero-projection edge [Figs.~\ref{fig:1}(a)(c)]. This bulk projection criterion precisely addresses the intricate geometry-dependent localization behaviors as shown in Figs.~\ref{fig:1}(a)-(c). 
It's important to note that for different OBC eigenvalues, the corresponding Fermi points and their projections on edges are generically changed. 
This may lead to different geometry-dependent localization behaviors for distinct OBC eigenvalues. 

The Hamiltonian $h(\bm{k})$ in Eq.(\ref{MT_Model}) respects reciprocity symmetry, which further restricts the spectral winding number. 
As shown in Fig.~\ref{fig:2}(a), despite $w(E_0,k_{y})$ being nonzero within a range of $k_y$, the reciprocity leads to a relation $w(E_0,k_{y})=-w(E_0,-k_{y})$, where red and blue colored regions correspond to the positive and negative projective winding numbers, respectively. 
Ultimately, the reciprocity symmetry enforces the total spectral winding number along every direction to be zero, expressed as 
\begin{equation}
    \forall \theta_i, \, \nu_{\theta_i}(E_0)=\int_{-\pi}^{\pi} dk_{\theta_i} w(E_0,k_{\theta_i})=0.
\end{equation} 
Here $w(E_0,k_{\theta_i})$ is the projective winding number for the edge along the $\theta_i$ direction, which is given by Eq.(\ref{MT_ProWinding}). 
Therefore, the symmetry guarantees that there is no net non-reciprocity along any spatial direction, which means that the wavefunction can be stably localized on the edges without being pushed into the corners. 
Our paper focuses on reciprocal systems that satisfy this condition of zero total spectral winding number (see Appendix~\ref{AppendixA}). 

\begin{figure*}[t]
    \begin{center}
    \includegraphics[width=1\linewidth]{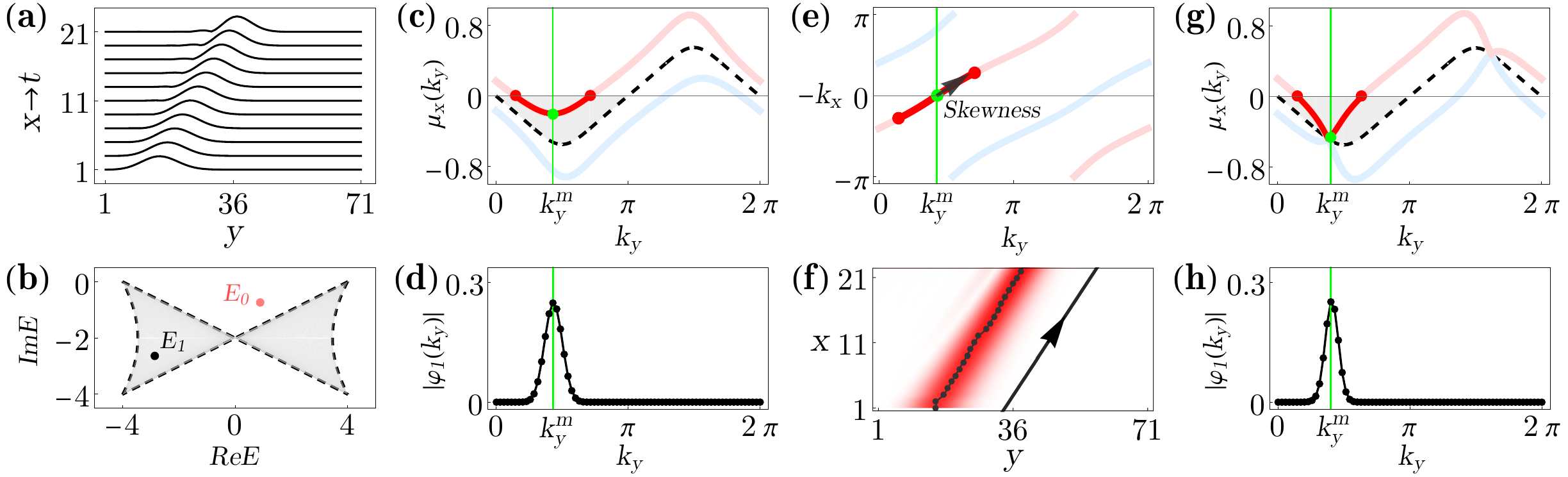}
    \par\end{center}
    \protect\caption{\label{fig:3}~Skewness of skin mode under a square geometry of system size $L_x=L_y=71$, with Hamiltonian parameters $t=2,\gamma=1$. 
    (a) illustrates the mode components in the first 20 $x$-layers, which have been renormalized in each layer. 
    (b) The OBC spectrum can be categorized into two regions: those outside and inside the gray region, respectively represented by $E_0 {=} 0.877 {-} 0.751 i$ and $E_1 {=} {-}2.867 {-} 2.656i$. 
    (c)-(f) gives the skewness of the skin mode with energy $E_0$. 
    (c) The solutions $\mu_{x,1}(k_y)$ (the blue line), $\mu_{x,2}(k_y)$ (the red line), and cylinder-geometry GBZ $\tilde{\mu}_x(k_y)$ (the black-dashed line). 
    The gray region represents the range between cylinder-geometry GBZ and BZ ($\mu_x(k_y)\equiv 0$). 
    The minimum in this range (the green dot) determines the maximum value of Fourier components of edge-layer mode components shown in (d). 
    Correspondingly, (e) indicates the skewness as the tangential direction of $-k_{x,2}(k_y)$ at $k_y^m$, denoted by the black arrow. 
    (f) compares the analytical results of skewness (the black-solid arrow) with the numerical results (the black dots similar to Fig.~\ref{fig:1}(e)). 
    (g)(h) correspond to the case of energy $E_1$. }
\end{figure*}

To better understand this bulk projection criterion, we map the non-Hermitian Hamiltonian $H_{\tx{OBC}}$ to a Hermitian chiral semimetal~\cite{BernevigBook}, 
\begin{equation}\label{MT_SemiMap}
    \tilde{H}_{\tx{OBC}}(E_0) = 
    \begin{pmatrix}
    0 & H_{\tx{OBC}} - E_0 \\ 
    H_{\tx{OBC}}^{\dagger}-E_0^{*} & 0 
    \end{pmatrix},
\end{equation}
with the identical open-boundary geometry. 
Under this mapping, the Fermi points of $E_0$ in non-Hermitian Hamiltonian map to band crossings, e.g., Dirac points, in the corresponding Hermitian semimetal; the projection region bounded by the Fermi points corresponds to the Fermi arc terminated at band crossings, as illustrated in Fig.~\ref{fig:2}(b). 
Most importantly, through this exact mapping, the edge-skin mode with energy $E_0$ corresponds to the zero-energy edge states of the Hermitian semimetal with fully OBCs, due to $\det[\tilde{H}_{\tx{OBC}}(E_0)] = |\det[H_{\tx{OBC}} - E_0]|^2=0$ (see more details in Appendix~\ref{AppendixB}). 
Here, the zero-energy edge states arise from the splitting of Fermi-arc states caused by fully OBCs [Fig.~\ref{fig:2}(b)]. 
This mapping conveys dual implications: 
(i) the bulk projection criterion in the non-Hermitian Hamiltonian is essentially inherited from the bulk-boundary correspondence in the corresponding Hermitian semimetal; 
(ii) conversely, the exact zero-energy edge states in Hermitian semimetals with fully OBC geometry, a topic largely overlooked despite an extensive research on Fermi-arc states~\footnote{Although Fermi-arc states are frequently discussed in Hermitian chiral semimetals with semi-infinite boundary conditions, fully OBCs, especially with specific geometries, are more relevant to realistic materials. Under fully OBCs, the degenerate Fermi-arc states become split, yet several zero-energy edge states remain. However, the characteristics of these zero modes are largely unexplored},
share the same spatial distribution and localization characteristics as the corresponding edge-skin modes, such as geometry-dependent localization and the skewness (Fig.~\ref{fig:1}). 

\section{Edge theory II: skewness of edge-skin modes}
To understand the skewness near the edge{-}$y$, as illustrated in Figs.~\ref{fig:1}(a)(d)(e), we treat the layer $x$ as time $t$, namely ${(x,y)\to (t,y)}$. 
From this perspective, for a given energy $E_0$, the 2D static wavefunction is mapped into a 1D space-time dynamics along the $y$ direction. 
For example, we consider the wavefunction shown in Fig.~\ref{fig:1}(a). 
Under the mapping $(x,y)\rightarrow (t,y)$, the near-edge distribution of skin mode, e.g., the first 20 $x$-layers as shown in Fig.~\ref{fig:1}(d), maps to a 1D finite-time wavepacket dynamics in the $y$ direction. 
When we renormalize the wavepacket at each time step (i.e., the wave component at each layer $x$), it becomes more clear that the wavepacket propagates with a velocity, as shown in Fig.~\ref{fig:3}(a). 
In this sense, the skewness, indicated by the black arrow in Fig.~\ref{fig:1}(e), is mapped into the wavepacket group velocity in the $y$ direction, as shown in Fig.~\ref{fig:3}(a). 
In the finite-time dynamics, the wavepacket has not yet encountered the boundaries in the $y$ direction [Fig.~\ref{fig:3}(a)], therefore, the velocity (or skewness) can be captured by the Hamiltonian with PBC in $y$~\cite{MaoLiang2021} and OBC in $x$ direction, effectively being in a cylinder geometry. 

We now present our key results of skewness using the example given by Eq.(\ref{MT_Model}).
The formula of skewness in generic cases is provided in Appendix~\ref{AppendixD}. 
With the effective cylinder geometry, $k_y$ is real-valued. We generalize $k_x$ into a complex value ${k_x \to k_x-i\mu_x}$ and take the substitution $\beta_x\coloneqq e^{ik_x+\mu_x}$. 
The corresponding non-Bloch Hamiltonian with real-valued edge momenta $k_y$ can be expressed as
\begin{equation}\label{MT_ModelCyl}
    \mathcal{H}(\beta_x, k_y) = c_{+}(k_y) \beta_x + c_{-}(k_y) \beta_x^{-1} - 2 i \gamma,
\end{equation}
where $c_{\pm}(k_y) = t e^{\pm i k_y}+i \gamma$. 
The solution of $\mu_x(k_y) \equiv \log|\beta_x(k_y)|$ can be obtained by solving $\det[\mathcal{H}(\beta_x,k_y)-E] = 0$. 
For a given $k_y$, the non-Bloch Hamiltonian reduces to a 1D subsystem. 
Therefore, the GBZ for the cylinder-geometry Hamiltonian is represented by
\begin{equation}\label{MT_GBZCyl}
    \tilde{\mu}_x(k_y) \equiv \mu_{x,1}(k_y,E) = \mu_{x,2}(k_y,E),
\end{equation}
which is satisfied only when $E$ is in the cylinder-geometry spectrum [denoted by $\sigma_{\tx{cyl}}$, the gray region in Fig.~\ref{fig:3}(b)]. 
Given that the OBC spectrum is generally inconsistent with the cylinder-geometry spectrum, the OBC eigenvalues can be sorted into two sections: $E \notin \sigma_{\tx{cyl}}$ and $E \in \sigma_{\tx{cyl}}$, as illustrated in Fig.~\ref{fig:3}(b). 

Under this space-time mapping, $x$ serves as the temporal parameter $t$, and its conjugate quantity, $-(k_x-i\mu_x)$, can be interpreted as the ``complex energy''. 
For a specified energy $E_0$, the $-(k_x-i\mu_x)$ is $k_y$ dependent according to the characteristic equation $\det[\mathcal{H}(\beta_x,k_y)-E_0] = 0$, where $\beta_x=e^{i(k_x-i\mu_x)}$. 
Therefore, for a fixed energy $E_0$, $-k_x(k_y)$ and $\mu_x(k_y)$ are regarded as the real and imaginary parts of the ``energy bands'' of $k_y$. 
In 1D non-Hermitian wavepacket dynamics~\cite{GongPRX2018}, the wavepacket group velocity is determined by the real part of the energy band, that is, $v(k_y)=-\partial_{k_y}k_x(k_y)$. 
Given this space-time mapping, the skewness of the wavefunction for energy $E_0$ is analogous to the group velocity, thereby being represented by $-\partial_{k_y}k_x(k_y)$. 
In 1D wavepacket dynamics, the group velocity is governed by the momentum center at the initial time; correspondingly, the skewness near the edge layer is determined by the momentum center of the mode component at the edge layer $x=1$. 
We denote the edge layer momentum center as $k_y^m$, as shown in Fig.~\ref{fig:3}(d). 
Now we state the first key conclusion of skewness for energy $E_0$: in the large-size limit, the edge momentum center $k_y^m$ is always determined by the minimum of $\mu_{x}(k_y^m,E_0)$ in the range $\tilde{\mu}_x(k_y) < \mu_x(k_y,E_0)<0$, as denoted by the green dot in Fig.~\ref{fig:3}(c). 

Here we provide a straightforward proof for this statement, with more details available in Appendix~\ref{AppendixC}. 
We begin with the wavefunction component at layer $x_0$, expressed as $\psi_{x_0}(y)$. The layer $x_0$ is assumed to be away from the edge layer and deep in the bulk. 
The numerical Fourier transform of the layer-$x_0$ component is given by $\varphi_{x_0}(k_y) = \frac{1}{\sqrt{L_y}}\sum\nolimits_{y=1}^{L_y} e^{-ik_y y} \, \psi_{x_0}(y)$. 
The $\psi_{x_0}(y)$ amplifies from the bulk to the edge along the $-x$ direction (or equivalently, decays in the $+x$ direction), the amplification rate for each $k_y$ Fourier component can be determined by the propagator
\begin{equation}\label{MT_KyProp}
    |\mathcal{G}_{E_0}(-x,k_y)| = \left| \oint_{C_{\beta_x}} \frac{d\beta_x}{\beta_x}\frac{\beta_x^{-x}}{E_0^+-\mathcal{H}(\beta_x,k_y)} \right| \propto e^{-\mu_x(k_y)x}, 
\end{equation}
where the integral contour $C_{\beta_x}$ is the cylinder-geometry GBZ and $E_0$ is the OBC eigenvalue of the skin mode. 
For a fixed $k_y$, the integral value is finally determined by the poles $\mu_x=\log|\beta_x|$ within the range between GBZ ($\tilde{\mu}_x(k_y)$, the black dashed line in Fig.~\ref{fig:3}(c)) and BZ [$\mu_x(k_y)\equiv 0$]. 
These contributing poles $\mu_x(k_y)$ are represented by the red opaque arc bounded by two Fermi points (indicated by the red dots) in the gray region in Fig.~\ref{fig:3}(c). 
Therefore, the minimum $\mu_x(k^m_y)$ in this range, denoted by the green dot in Fig.~\ref{fig:3}(c), leads to the maximal amplification rate in Eq.(\ref{MT_KyProp}) and finally determines the momentum center at the edge layer $x=1$, as shown by a numerical verification in Fig.~\ref{fig:3}(d). 

In the case of skin mode with energy $E_0\notin \sigma_{\tx{cyl}}$ [Fig.~\ref{fig:3}(b)], there is always a gap between two different imaginary parts $\mu_{x,1}(k_y)< \tilde{\mu}_x(k_y) < \mu_{x,2}(k_y) < 0$ [Fig.~\ref{fig:3}(c)]. 
The cylinder-geometry GBZ $\tilde{\mu}_x(k_y)$ resides in this gap, as indicated by the black-dashed line in Fig.~\ref{fig:3}(c). 
The `gap' here specifically means that $\mu_{x,i}(k_y)$ does not intersect with the cylinder-geometry GBZ $\tilde{\mu}_x(k_y)$. 
Based on our first conclusion, the edge momentum center $k_y^m$ always corresponds to the minimum of $\mu_{x,2}(k_y)$, as illustrated in Figs.~\ref{fig:3}(c) and (d). 
In this case, it further satisfies: 
\begin{equation}\label{MT_StableCondition}
    \partial_{k_y} \mu_{x,2}(k_y^m) = 0
\end{equation}
It allows that the group velocity of the initial wavepacket does not change quickly, and the wavepacket can propagate at a constant velocity within a finite time. 
This wavepacket group velocity, exactly the skewness, is given by: 
\begin{equation}\label{MT_Skewness}
    s_{\tx{edge{-}y}}(E_0) = - \partial_{k_y}k_{x,2}(k_y^m),
\end{equation}
which is denoted by the black arrow in Fig.~\ref{fig:3}(e).
As an example, we select the skin mode with energy $E_0{=}0.877 {-} 0.751 i$. 
By solving the equation $\det[\mathcal{H}(\beta_x,k_y)-E_0]=0$, we can obtain the solutions $\mu_x(k_y)$. The edge layer momentum center $k_y^m$ can be determined by finding its minimum using Eq.(\ref{MT_StableCondition}). 
Finally, the skewness can be analytically calculated according to Eq.(\ref{MT_Skewness}) as $s_{\tx{edge{-}y}}(E_0) {=} 1.113$. 
In Fig.~\ref{fig:3}(f), we plot the analytical result $y(x)= s_{\tx{edge{-}y}}(E_0) \, x + y_0$ as the black arrow, and compare it with the numerical results (the black dots similar to Fig.~1(e)), which shows perfect agreement. 
Therefore, we demonstrate that the skewness of the skin mode is only determined by the non-Bloch Hamiltonian with real-valued edge momenta without requiring any boundary details, thus serving as an intrinsic localization quantity of the edge-skin modes.  

For the case of energy $E_1\in \sigma_{\tx{cyl}}$, the corresponding solutions $\mu_{x,i}(k_y)$ must intersect with $\tilde{\mu}_{x}(k_y)$ [Fig.~\ref{fig:3}(g)]. 
The edge-layer momentum distribution is depicted in Fig.~\ref{fig:3}(f). 
However, in this case the edge momentum center may not satisfy Eq.(\ref{MT_StableCondition}). 
Therefore, the skewness of skin mode will bend quickly when it moves away from the edge layer (see discussions in Appendix~\ref{AppendixE}). 
These findings on skewness can be directly generalized and applicable to more generic models with long-range hopping, as presented in Appendix~\ref{AppendixD}. 

\section{Anomalous spectral sensitivity in the edge-skin effect}

\begin{figure}[t]
    \begin{center}
    \includegraphics[width=1\linewidth]{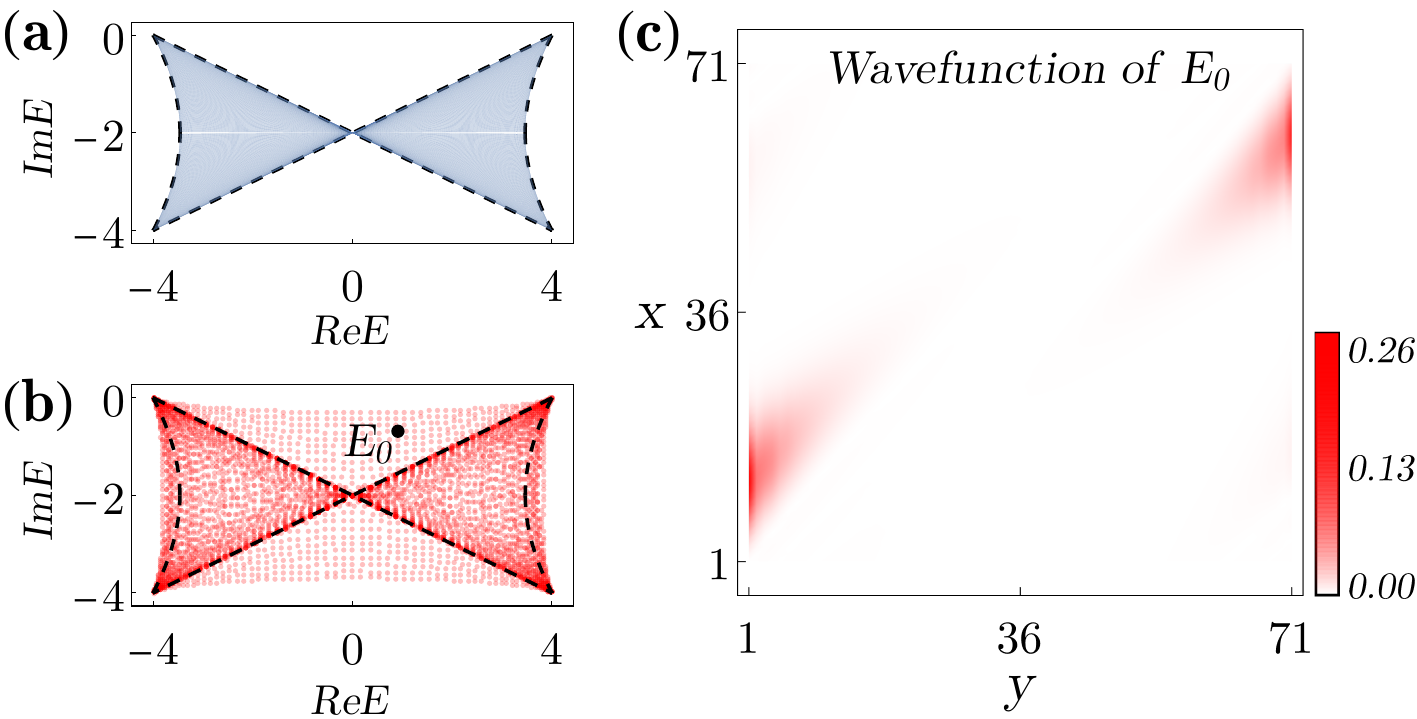}
    \par\end{center}\protect\caption{\label{fig:4}~Introducing weak local random disturbances with strength $W = 0.1$, as described in Eq.(\ref{eq:OnSiteDis}), into the cylinder-geometry Hamiltonian leads to a remarkable shift in the spectrum from the blue region in (a) to the red dots in (b). 
    (c) shows the spatial distribution of the corresponding wavefunction with energy $E_0 = 0.89 - 0.7i$, as indicated by the black dot in (b).} 
\end{figure}

The reciprocity symmetry in the bulk Hamiltonian described by Eq.(\ref{MT_ModelCyl}) ensures that the characteristic equation satisfies
\begin{equation}
\begin{split}
f_E(k_x,\mu_x,k_y) = f_E(-k_x,-\mu_x,-k_y) = 0,
\end{split}
\end{equation}
which we term fragile reciprocity. 
Note that fragile reciprocity is readily violated by factors such as onsite disorders or open boundaries that break translation in $y$ direction, hence the name of ``fragile''.
With this symmetry, the non-Bloch Hamiltonian in Eq.(\ref{MT_ModelCyl}) has two key features: (i) the $k_y$ components are decoupled with each other, and (ii) the doubly degenerate non-Bloch waves with opposite $k_y$ exhibit opposite localization in the $x$ direction. 
When the translation in $y$ direction is broken, these non-Bloch waves couple with each other and reconstruct the eigen wavefunction. 
The different range of $k_y$ contributes to different edge localization. 
For example, as shown in Fig.~\ref{fig:3}(c), the red opaque arc contributes to the localization near the edge layer $x=1$, which leads to the directional decay direction away from this edge, with the decay direction given by the skewness in Eq.(\ref{MT_Skewness}). 
Therefore, the skewness of the associated wavefunctions can be observed when the fragile reciprocity is broken. 
Meanwhile, a remarkable sensitivity of the spectrum can be expected, due to the breakdown of edge momentum conservation. 

Numerically, we introduce a line of weak onsite disorders into the cylinder Hamiltonian Eq.(\ref{MT_ModelCyl}) to break the fragile reciprocity, 
\begin{equation}\label{eq:OnSiteDis}
    \hat{V} = \sum\nolimits_{x} V_{x} \, \hat{c}^{\dagger}_{x,y_0}\hat{c}_{x,y_0}; \,\,\,\, V_{x} \in [0,W],
\end{equation}
where $V_{x}$ represents the strength of onsite potential at the lattice site $(x,y_0)$ with fixed $y=y_0$ and is randomly distributed in the range $[0, W]$. 
As shown in Fig.~\ref{fig:4}(a) and (b), the spectrum is remarkably shifted from the blue region in (a) to the red dots in (b) upon introducing weak onsite disturbances with a strength of $W=0.1$. 
Accordingly, as shown in Fig.~\ref{fig:4}(c), the spatial distribution of wavefunction with energy $E_0$ selected in Fig.~\ref{fig:4}(b) shows the skewness. 

We emphasize that in the case of the corner-skin effect or 1D non-Hermitian skin effect, the weak onsite disorders can be typically treated as perturbations to the spectrum~\cite{Kai2023PRB}; see more examples in Appendix~\ref{AppendixF}. 
Therefore, this anomalous spectral sensitivity and the emergence of skewness are unique to the edge-skin effect and crucially distinguished from the corner-skin effect. 

\section{Conclusion}
In summary, we establish an effective edge theory to describe the higher-dimensional edge-skin modes. 
For a generic edge-skin mode, the edge theory includes two parts: (i) we propose a bulk projection criterion to diagnose the localized edge; (ii) we find an intrinsic decay direction of the skin mode, termed skewness, which can be accurately captured by the non-Bloch bulk Hamiltonian with real-valued edge momenta without requiring any boundary details. 
Lastly, unique to the edge-skin effect, we reveal the anomalous sensitivity of the spectrum to perturbations that break fragile reciprocity, accompanied by the emergence of the skewness in the corresponding wavefunctions. 

\section*{Acknowledgments}

The authors thank Siddhartha Sarkar for his valuable discussions. 
This work was supported in part by the Office of Naval Research MURI Grant No. N00014-20-1-2479 (K.Z. and K.S.), Gordon and Betty Moore Foundation Award GBMF10694 (K.S.), and the Office Navy Research Award No. N00014-211-2770 (K.S.). 
Z.Y. was sponsored by the National Key Research and Development Program of China (Grant No. 2023YFA1407500), the National Natural Science Foundation of China (12322405, 12104450, 12047503), the Fellowship of China National Postdoctoral Program for Innovative Talents (BX2021300), the Fellowship of China Postdoctoral Science Foundation (2022M713108), and the Fundamental Research Funds for the Central Universities (20720230011). 

\appendix

\section{Reciprocity in the edge skin effect}\label{AppendixA}

This section is divided into two parts. First, we prove that the existence of Fermi points for open-boundary eigenvalues is guaranteed by the reciprocity of the Hamiltonian. Then, we discuss the relationship between reciprocity, the localization property of skin modes, and the zero total spectral winding number.

\subsection{Fermi points ensured by reciprocity}

For a given Bloch Hamiltonian $h(k_x,k_y)$, we first define the spectral winding number in the $x$ direction with fixed $k_y$
\begin{equation}\label{SM_SpecWinding1D}
    w(E_0, k_y) = \frac{1}{2\pi} \int_{-\pi}^{\pi} d k_x \, \partial_{k_x} \arg{\det{[h(k_x,k_y)-E_0]}},
\end{equation}
where $E_0$ represents a generic OBC eigenvalue, serving as a reference energy here. 

The reciprocity of the Hamiltonian ensures that the OBC bulk eigenvalues must be included in its PBC spectrum. Here, we provide a proof for this statement in single-band cases. 
Suppose an open-boundary bulk eigenvalue $E_0 \notin \sigma_{\text{PBC}}$, and the Hamiltonian respects reciprocity. 
It is known that the OBC bulk spectrum is obtained from the collapse of the PBC spectrum. Thus, the OBC bulk spectrum cannot exceed the outer boundary of the PBC spectrum. 
If we assume $E_0 \notin \sigma_{\text{PBC}}$, the only possibility is that $E_0$ is located within an inner gap of the PBC spectrum, such as some inner holes of the PBC continuum spectrum. 
Below, we prove that this case is forbidden by the reciprocity of the Hamiltonian. Consequently, open-boundary eigenvalues must belong to the PBC spectrum of $h(k_x,k_y)$, and for a given open-boundary eigenvalue $E_0$, there exist Fermi points ensured by reciprocity. 

Due to the reciprocity of the Hamiltonian, the spectral winding number in Eq.(\ref{SM_SpecWinding1D}) satisfies $w(E_0, k_y)=-w(E_0, -k_y)$, as detailed in Eq.(\ref{SecI_SymmConstraint}) below. 
At the high symmetry points ${k^{\ast}_y=0,\pi}$, we have $k^{\ast}_y=-k^{\ast}_y$. Therefore, for a generic OBC eigenvalue $E_0$, the spectral winding number satisfies $w(E_0, k^{\ast}_y)=-w(E_0, k^{\ast}_y)$. 
It implies that the PBC spectra of $h(k_x,k_y^{\ast})$ collapse into arcs in the complex-energy plane~\cite{Kai2020PRL}. 
Similarly, the spectra of $h(k^{\ast}_x,k_y)$ also collapse into arcs at the high symmetry points ${k_x^{\ast}=0,\pi}$. 
Assuming $E_0$ is in the inner gap of the PBC spectrum, there is a $k_y^0$ such that the spectrum of $h(k_x,k_y^0)$ has a nonzero spectral winding number with respect to $E_0$. 
When we smoothly transition $k_y$ from $k_y^0$ to $k_y^{\ast}$ (for example, $k^{\ast}_y=0$), the corresponding PBC spectrum of the 1D subsystem of $k_x$ must transition smoothly since the single-band Hamiltonian $h(k_x,k_y)$ is a holomorphic function of $k_x$ and $k_y$. 
Therefore, it is impossible to squeeze the spectrum from a loop encircling energy $E_0$ (at $k_y=k_y^0$) into an arc (at $k_y=k_y^{\ast}$) without passing through $E_0$. 
Consequently, the PBC spectrum must sweep through $E_0$ at some $k_{y,i}$. These $k_{y,i}$ correspond to the Fermi points of $E_0$. Therefore, open-boundary eigenvalue $E_0$ must belong to the periodic-boundary spectrum of $h(k_x,k_y)$. 

\subsection{Reciprocity, localization property, and the total spectral winding number}

Here, we classify the (non-Hermitian) skin modes in two-dimensional systems according to the total spectral winding number. 
The total spectral winding number for the $y$ direction is defined as the integral of $w(E_0, k_y)$ in Eq.(\ref{SM_SpecWinding1D}) over $k_y$: 
\begin{equation}
    \nu_{y}(E_0) = \frac{1}{2\pi} \int_{-\pi}^{\pi} d k_y \, w(E_0, k_y), 
\end{equation}
which depends on the reference energy $E_0$ and direction $y$. 
Based on the above definition, we have the total spectral winding for an arbitrary direction $\theta$, mathematically expressed as, 
\begin{equation}
    \begin{split}
    &\nu_{\theta}(E_0) = \frac{1}{2\pi} \int_{-\pi}^{\pi} d k_\theta \, w(E_0, k_\theta) \\
    &= \frac{1}{(2\pi)^2} \int_{-\pi}^{\pi} d k_\theta \, \int_{-\pi}^{\pi} d k_\bot \partial_{k_{\bot}} \arg{\det{[h(k_{\bot},k_\theta)-E_0]}}, 
    \end{split}
\end{equation}
where $k_{\bot}$ represents the momentum along the normal direction of the $\theta$ direction. 
Mathematically, there are two mutually exclusive and complete cases: 
\begin{equation}\label{SM_SkinClass}
	\begin{split}
		&(\mathrm{i}) \,\,\, \exists \, \theta; \, \nu_{\theta}(E_0)\neq 0; \\
		&(\mathrm{ii}) \,\, \forall \, \theta; \, \nu_{\theta}(E_0) = 0.
	\end{split}
\end{equation}
These two cases correspond to two classes of skin modes: the corner-skin modes in case (i) and the edge-skin modes in case (ii). 
It is important to note that case (ii) can be enforced by reciprocity in the bulk Hamiltonian, which we will discuss later. 

In case (i), the skin mode prefers to localize at a specific corner of a polygon, such as a square geometry, thus termed corner-skin mode. 
As an illustration, consider a skin mode with energy $E_0$ in a square geometry under case (i), generally, $\nu_{x}(E_0)\neq 0$ and $\nu_{y}(E_0)\neq 0$. 
The nonzero total spectral winding implies $w(E_0, k_x)\neq 0$ for some $k_x$ and $w(E_0, k_y)\neq 0$ for some $k_y$ as defined in Eq.(\ref{SM_SpecWinding1D}). 
We generalize $k_x$ and $k_y$ into complex value $k_x - i \mu_x$ and $k_y - i \mu_y$, respectively. 
According to the recently established amoeba formula~\cite{HYWang2022,HPHu2023}, one can always find suitable $\mu_x\neq 0$ and $\mu_y\neq 0$ such that the total spectral winding number vanishes:
\begin{equation*}
    \begin{split}
    &\tilde{\nu}_x(E_0) = \frac{1}{(2\pi)} \int_{-\pi}^{\pi} d k_x \, w(E_0,k_x-i\mu_x) = 0 ; \\ 
    &\tilde{\nu}_y(E_0) = \frac{1}{(2\pi)} \int_{-\pi}^{\pi} d k_y \, w(E_0,k_y-i\mu_y) = 0 .
    \end{split}
\end{equation*}
The nonzero $\mu_x$ and $\mu_y$ give us the effective localization length in $x$ and $y$ directions, respectively. 
It means that the corresponding skin mode will localize at the corner of a square lattice. 

\begin{figure*}[t]
    \begin{center}
        \includegraphics[width=\linewidth]{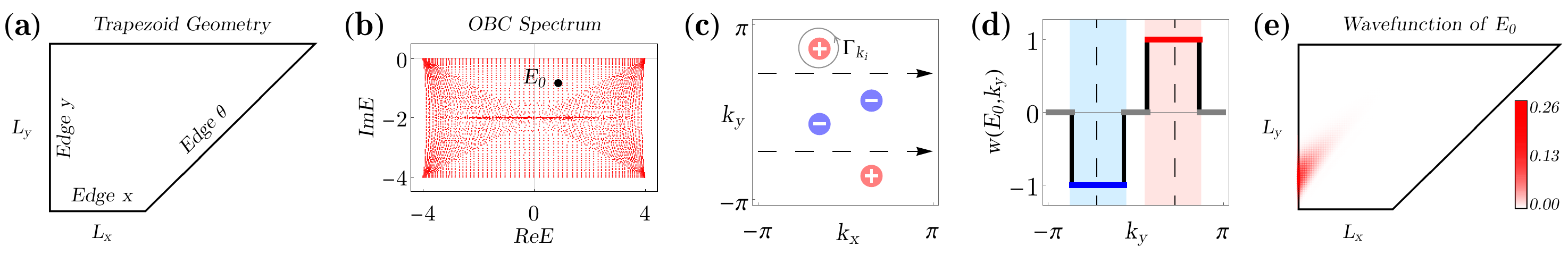}
        \par\end{center}
    \protect\caption{\label{fig:S1} 
    The illustration for the bulk-edge correspondence in the edge-skin effect. 
    (a) The lattice geometry is depicted with a trapezoid of size $L_x=45$, $L_y=71$, and the edge $\theta$ oriented in the direction of $\theta=\pi/4$. 
    (b) The eigenvalues of the Hamiltonian in Eq.(\ref{SM_EdgeSkinModel}) under the trapezoid geometry are depicted by the red dots. 
    A specific OBC eigenvalue, $E_0=0.866-0.814i$, is indicated by the black dot. 
    The Fermi points of $E_0$ are shown in (c), where the sign `$\pm$' represents the topological charge of each Fermi point. 
    We select two paths with fixed $k_y$, indicated by the black arrows, the corresponding spectral winding numbers are shown in (d). $w(E_0,k_y)=\pm 1$ for a range of $k_y$. According to the projection criterion, the skin mode with energy $E_0$ is localized on the edge $y$, as illustrated in (e).
    The system parameters are set to $t=2,\gamma=1$. } 
\end{figure*}

In fact, the two-dimensional corner-skin effect is very similar to the one-dimensional skin effect. 
In one dimension, case (i) defined by Eq.(\ref{SM_SkinClass}) reduces to a nonzero Bloch spectral winding number, namely, $\nu(E_0) = \frac{1}{(2\pi)} \int_{-\pi}^{\pi} dk \, \partial_{k}\mathrm{arg}\,\mathrm{det}[h(k)-E_0 ]\neq 0$.
It was known that in one dimension, nonzero Bloch spectral winding number regarding the reference energy $E_0$ means that the OBC wavefunction of $E_0$ appears as a skin mode under OBCs~\cite{Kai2020PRL,Okuma2020_PRL}. 
Similarly, one can extend $k$ into complex value $k-i\mu$ such that the spectral winding number vanishes, that is, $\tilde{\nu}(E_0) = \frac{1}{(2\pi)} \int_{-\pi}^{\pi} dk \, \partial_{k}\mathrm{arg}\,\mathrm{det}[h(k-i\mu)-E_0 ] = 0$. 
Finally, nonzero $\mu$ characterizes the localization length of the skin mode with energy $E_0$. 

Obviously, the above arguments do not apply to case (ii), since the total spectral winding is always zero. 
It implies that the localization feature in case (ii) is essentially different from the corner-skin effect in case (i). 
In case (ii), the zero total spectral winding means that there is no net non-reciprocity along any spatial direction. 
The skin modes in case (ii) are generally localized on the edges of a polygon, thus termed edge-skin effect, which is our focus. 

Here we show that the reciprocity in the bulk Hamiltonian $h(\bm{k})$ can guarantee case (ii) in Eq.(\ref{SM_SkinClass}), thereby enforcing the Hamiltonian to manifest the edge-skin effect. 
The Hamiltonian that respects the reciprocity can be expressed as: 
\begin{equation}\label{SM_SymmEdgeSkin}
    \begin{split}
    \mathcal{U}h^{T}(\bm{k})\mathcal{U}^{\dagger} = h (-\bm{k}).
    \end{split}
\end{equation}
Here, $\mathcal{U}$ represents the unitary representation part of the reciprocity and satisfies $U^T=U$. 
With these symmetries, the spectral winding number defined in Eq.(\ref{SM_SpecWinding1D}) satisfies: 
\begin{equation}\label{SecI_SymmConstraint}
    \begin{split}
        w(E_0,k_{y}) & = \frac{1}{2\pi} \int_{-\pi}^{\pi} dk_{x}  \, \partial_{k_{x}}\arg{\det{[h(k_x,k_y)-E_0]}} \\
        & = \frac{1}{2\pi} \int_{-\pi}^{\pi} dk_{x}  \, \partial_{-k_{x}}\arg{\det{[h(-k_x,k_y)-E_0]}} \\
        &= - \frac{1}{2\pi} \int_{-\pi}^{\pi} dk_{x}  \, \partial_{k_{x}}\arg{\det{[h(k_x,-k_y)-E_0]}} \\ & = - w(E_0,-k_{y}),
    \end{split}
\end{equation}
where the third equality holds due to the reciprocity in the bulk Hamiltonian, as shown in Eq.~(\ref{SM_SymmEdgeSkin}). 
Therefore, with this symmetry, the total spectral winding number over $k_y$ must be zero, expressed as: 
\begin{equation}\label{SM_ZeroYWinding}
    \nu_{y}(E_0) = \frac{1}{2\pi} \int_{-\pi}^{\pi} d k_y w(E_0,k_y) = 0. 
\end{equation}
In fact, the reciprocity necessitates zero total spectral winding along every direction, namely, 
\begin{equation}\label{SM_EdgeSkinCond}
    \forall \theta; \,\,\,\, \nu_{\theta}(E_0) = 0,
\end{equation}
which corresponds to case (ii) defined in Eq.(\ref{SM_SkinClass}). 
In conclusion, reciprocity ensures that the OBC spectrum is always included by PBC spectrum and enforces zero total spectral winding number. 
Therefore, the non-Hermitian Hamiltonian respecting reciprocity exhibits the edge skin effect. 
One example is the Hamiltonian given by Eq.(1) in the main text. 
The single-band model Hamiltonian respects reciprocity, therefore, it exhibits edge-skin effect with fully OBCs. 

Notably, in one dimension, reciprocity symmetry requires the spectral winding number to vanish and hence forbids the non-Hermitian skin effect~\cite{Kawabata2020,YYFPRL2020}. 
In sharp contrast, in two and higher dimensions, the edge-skin effect can coexist with reciprocity. 
Therefore, edge-skin effect is unique to higher dimensions. 

\section{The bulk projection criterion and mapping to Hermitian semimetals}\label{AppendixB}

Here, we elaborate on the bulk projection criterion and its mapping to Hermitian semimetals by presenting a numerical example. 
For a given edge-skin mode with energy $E_0$, the corresponding bulk projection criterion determines the localized edges. 
On one hand, the exact mapping helps to understand why the bulk projection criterion works well. 
On the other hand, through this exact mapping, the localization features of the edge-skin mode, such as geometry-dependent localization and the skewness, can be directly applied to the zero-energy edge states in the Hermitian semimetal with fully OBCs. 
Notably, although the Fermi-arc states in semimetals under semi-infinite boundary conditions are frequently investigated~\cite{BernevigBook}, the exact zero-energy edge states under fully OBCs are often overlooked. 

\subsection{The bulk projection criterion}

We use the example of edge-skin effect discussed in the main text. 
The Bloch Hamiltonian is given by
\begin{equation}\label{SM_EdgeSkinModel}
	h(\bm{k}) = 2 t \cos{(k_x+k_y)} - 2 i \gamma (1-\cos{k_x}),
\end{equation}
with parameters $t=2, \gamma=1$ set for our discussion. 
Under the open boundary condition with a trapezoid geometry, as illustrated in Fig.~\ref{fig:S1}(a), the OBC eigenvalues can be evaluated and represented by red dots in Fig.~\ref{fig:S1}(b). 
Here, the system size is $L_y=71, L_x=41$. 
For a generic OBC eigenvalue $E_0=0.866 - 0.814 i$, indicated by the black dot in Fig.~\ref{fig:S1}(b), we have a set of Fermi points $\bm{k}_i$ that satisfy $\det{[h(\bm{k}_i)-E_0]}=0$, as shown in Fig.~\ref{fig:S1}(c).
For each Fermi point, we can assign a topological charge that is defined as:
\begin{equation}
    C_{E_0}(\bm{k}_i) = \frac{1}{2\pi}\oint_{\Gamma_{\bm{k}_i}} d\bm{k} \, \partial_{\bm{k}}\arg{\det[h(\bm{k}_i)-E_0]},
\end{equation}
where $\Gamma_{\bm{k}_i}$ represents the integral contour encircling Fermi point $\bm{k}_i$ in a counterclockwise manner. 
In such a way, the topological charges of these four Fermi points of $E_0$ can be obtained and indicated by the `$\pm$' signs corresponding to $C_{E_0}(\bm{k}_i)=\pm 1$. 
As shown in Fig.~2(a) of the main text (the same as that in Fig.~\ref{fig:S1}(c)), these four Fermi points have zero projection onto edge-$x$ and edge-$\theta$, while they have nonzero projection onto edge-$y$.
It means that $w(E_0,k_y)\neq 0$ given by Eq.(\ref{SM_SpecWinding1D}) for a range of $k_y$, as illustrated in Fig.~\ref{fig:S1}(d). 
Meanwhile, due to the reciprocity symmetry in the Hamiltonian Eq.(\ref{SM_EdgeSkinModel}), the spectral winding satisfies $w(E_0,k_y)=-w(E_0,-k_y)$, as shown in Fig.~\ref{fig:S1}(d), which ensures the zero total spectral winding number (Eq.(\ref{SM_EdgeSkinCond})). 

\begin{figure*}[t]
    \begin{center}
        \includegraphics[width=1\linewidth]{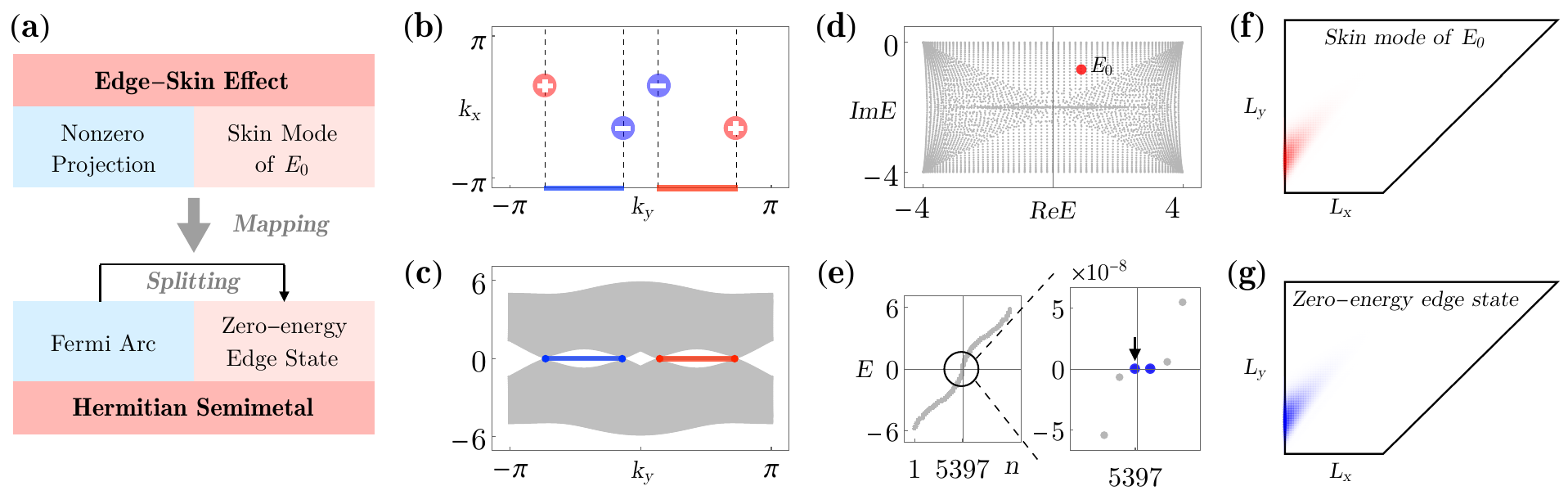}
        \par\end{center}
    \protect\caption{\label{fig:S2}
    (a) Mapping from the non-Hermitian edge-skin effect to a Hermitian semimetal is detailed. 
    Specifically, the Fermi points that have nonzero projections on the edge $y$ in (b) are mapped to Dirac points featuring connected Fermi-arcs in (c). 
    In (c), the energy bands are calculated using the Hamiltonian Eq.(\ref{SecI_BlochMapping}) with $E_0=0.866-0.814i$, under PBC in the $y$ and OBC in the $x$ direction, and with $L_x=61$.
    The OBC eigenvalue $E_0$ (d) in the non-Hermitian Hamiltonian Eq.(\ref{SM_EdgeSkinModel}) is mapped to two exact zero energies (e) in the Hermitian semimetal Eq.(\ref{SecI_BlochMapping}). 
    In (e), the horizontal axis $n$ represents the $n$-th eigenvalue ordered by their magnitudes. 
    Correspondingly, the skin mode of $E_0$ (f) is mapped into the zero-energy edge state (g). They show the same spatial distribution under the same lattice geometry. 
    Notably, the two zero-energy edge states remain after the splitting of Fermi-arc states induced by the fully OBCs. 
    The system parameters are set to $t=2,\gamma=1$. 
    } 
\end{figure*}

According to the bulk projection criterion, the wavefunction of $E_0$ does not exhibit localization on the open-boundary geometry that is composed of edge-$x$ and edge-$\theta$ (as shown in Fig.1(b) in the main text), whereas it localizes at edge-$y$ in the trapezoidal geometry, as depicted in Fig.~\ref{fig:S1}(e).
It is important to note that with different OBC eigenvalues $E_0$, the distribution of Fermi points in the BZ will be different. 
Consequently, the projection of Fermi points on different edges changes. 
It may lead to distinct localization behaviors of the corresponding wavefunction. 

\subsection{The mapping to a Hermitian semimetal}

Here we utilize an exact mapping that maps the non-Hermitian Hamiltonian with the edge-skin effect to a Hermitian chiral semimetal, which helps to better understand the bulk projection criterion. 

Under this mapping~\cite{GongPRX2018,SatoPRX2019}, the non-Hermitian Bloch Hamiltonian in Eq.(\ref{SM_EdgeSkinModel}) is mapped into 
\begin{equation}\label{SecI_BlochMapping}
    \begin{split}
        &\tilde{h}_{E_0}(\bm{k}) = \begin{pmatrix}
            0 & h(\bm{k}) - E_0 \\ 
            h^{\dagger}(\bm{k}) - E_0^{\ast} & 0 
        \end{pmatrix} \\
        & = \left({2 t \cos(k_x+k_y) - E_0^r}\right) \sigma_x {+} 
        \left({2\gamma (1-\cos{k_x})+E_0^i}\right) \sigma_y,
    \end{split}
\end{equation}
where $E_0^r, E_0^i$ represent the real and imaginary parts of the complex $E_0$, respectively, and $\sigma_x$ and $\sigma_y$ are the Pauli matrices. 
It should be noted that the Hermitian Hamiltonian always respects an additional chiral symmetry,
\begin{equation*}
    \sigma_z \tilde{h}_{E_0}(\bm{k})\sigma_z = - \tilde{h}_{E_0}(\bm{k}).
\end{equation*}
Since $\det[h(\bm{k})-E_0] = 0$ at Fermi points, the mapped Hermitian Hamiltonian satisfies 
\begin{equation*}
    \det[\tilde{h}_{E_0}(\bm{k})] = {\left|\det[h(\bm{k})-E_0]\right|}^2 = 0
\end{equation*}
at these $\bm{k}$ points, that is, Dirac points in Hermitian semimetal, as shown in Fig.~\ref{fig:S2}(c). 
It means that the Hermitian Hamiltonian $\tilde{h}_{E_0}(\bm{k})$ is always gapless at zero energy and respects the chiral symmetry. 
Therefore, the mapped Hermitian Hamiltonian is always a chiral semimetal~\cite{BernevigBook}. 

Through this mapping, the Fermi points in the non-Hermitian Hamiltonian map to Dirac points in the Hermitian chiral semimetal. 
The nonzero projection region of Fermi points in non-Hermitian Hamiltonian corresponds to the Fermi arc in the Hermitian semimetal, as illustrated in Fig.~\ref{fig:S2}(b)(c). 

Now we show that the non-Hermitian edge-skin mode with $E_0$ can be mapped into the zero-energy edge states in the Hermitian semimetal. 
The mapping can be formally expressed as: 
\begin{equation}\label{SM_OBCMapping}
    \tilde{H}_{\tx{OBC}}(E_0) = 
    \begin{pmatrix}
    0 & H_{\tx{OBC}} - E_0 \\ 
    H_{\tx{OBC}}^{\dagger}-E_0^{*} & 0 
    \end{pmatrix},
\end{equation}
with the same open-boundary geometry, for example, the trapezoid geometry depicted in Fig.~\ref{fig:S1}(a). 
Under the fully OBCs, the Fermi-arc states with order-$L$ ($L$ generally representing the lattice length) will split. 
However, there are always the order-1 zero-energy edge states survived, due to $\det[\tilde{H}_{\tx{OBC}}(E_0)] = \left|\det[H_{\tx{OBC}} - E_0]\right|^2=0$. 
Therefore, the complex energy $E_0$ in the non-Hermitian Hamiltonian, as indicated by the red dot in Fig.~\ref{fig:S2}(d) is mapped into the zero-energy edge states, denoted by the blue dots in Fig.~\ref{fig:S2}(e). 

For a given energy $E_0$ in the non-Hermitian Hamiltonian, we can define the left and right eigenvectors as follows,
\begin{equation}\label{SecI_NHBiBasis}
    H_{\tx{OBC}}|\psi^{R}_{E_0}\rangle = E_0|\psi^{R}_{E_0}\rangle; \quad 
    H_{\tx{OBC}}^{\dagger}|\psi^{L}_{E_0}\rangle = E^{\ast}_0 |\psi^{L}_{E_0}\rangle.
\end{equation}
Here the superscripts $R/L$ denote the right/left eigenvectors. 
The skin mode with energy $E_0$ refers to the right eigenvector, $\psi_{E_0}(\bm{r}) \equiv \langle \bm{r} | \psi^{R}_{E_0}\rangle$. 
Using the mapping given by Eq.(\ref{SM_OBCMapping}), the Hermitian Hamiltonian with the same OBC geometry must have two degenerate zero-energy edge states, expressed as 
\begin{equation}
    \begin{split}
        &\tilde{H}_{\tx{OBC}}(E_0)|\tilde\psi_0\rangle = \tilde{H}_{\tx{OBC}}(E_0) 
        \begin{pmatrix}
        \bm{0} \\ |\psi^R_{E_0}\rangle 
        \end{pmatrix} = 0; \\          
        &\tilde{H}_{\tx{OBC}}(E_0)|\tilde\psi^{\prime}_0\rangle =
        \tilde{H}_{\tx{OBC}}(E_0) 
        \begin{pmatrix}
        |\psi_{E_0}^{L}\rangle  \\ \bm{0}
        \end{pmatrix} = 0,
    \end{split}
\end{equation}
where $\bm{0}$ denotes the zero column vector. 
We emphasize that the zero-energy edge states $|\tilde{\psi}_0\rangle$ and $|\tilde{\psi}^{\prime}_0\rangle$ under OBCs arise from the splitting of Fermi-arc states. 
There are two degrees of freedom per unit cell, we define the spatial distribution of zero-energy edge states as 
\begin{equation}
    \tilde{W}_0(\bm{r}) = \sum_{\alpha=A,B} |\tilde{\psi}_{0,\alpha}(\bm{r})|^2; \quad 
    \tilde{W}^{\prime}_0(\bm{r}) = \sum_{\alpha=A,B} |\tilde{\psi}^{\prime}_{0,\alpha}(\bm{r})|^2.
\end{equation}
Here $\tilde{\psi}_{0,A}(\bm{r})=0$ and $\tilde{\psi}_{0,B}(\bm{r})=\psi^R_{E_0}(\bm{r})$, $\tilde{\psi}^{\prime}_{0,A}(\bm{r})=\psi^L_{E_0}(\bm{r})$ and $\tilde{\psi}^{\prime}_{0,B}(\bm{r})=0$. 

In the edge-skin effect, the reciprocity requires the Hamiltonian matrix in real space to satisfy $H^{T}=H$. 
It follows from Eq.(\ref{SecI_NHBiBasis}) that 
\begin{equation*}
    (H^{\dagger}-E_0^{\ast})|\psi_{E_0}^L\rangle = (H^{T}-E_0)|\psi_{E_0}^L\rangle^{\ast} = (H-E_0)|\psi_{E_0}^L\rangle^{\ast} = 0.
\end{equation*}
Comparing this with the eigenequation for the right eigenvector, we obtain
$$|\psi^R_{E_0}(\bm{r})|^2 = |\psi^L_{E_0}(\bm{r})|^2.$$
Finally, we conclude that the two degenerate zero modes and the edge-skin mode $|\psi^R_{E_0}\rangle$ share identical spatial distributions, expressed as
\begin{equation}
    \tilde{W}_0(\bm{r})  =\tilde{W}^{\prime}_0(\bm{r}) = W_{E_0}(\bm{r}) = |\psi_{E_0}^R(\bm{r})|^2. 
\end{equation}
As an example, we select $E_0=0.866-0.814 i$ as denoted in Fig.~\ref{fig:S2}(d). 
The corresponding skin mode under the trapezoid geometry (Fig.~\ref{fig:S1}(a)) is plotted Fig.~\ref{fig:S2}(f), with system size $L_y=71$ and $L_x =41$. 
Through the mapping defined in Eq.(\ref{SM_OBCMapping}), the spatial distribution of the zero-energy edge state indicated by the black arrow in Fig.~\ref{fig:S2}(e) is depicted in Fig.~\ref{fig:S2}(g). 
They show the identical spatial distribution. 
In conclusion, the mapping between non-Hermitian edge-skin effect and Hermitian semimetal is illustrated in Fig.~\ref{fig:S2}(a). 

\section{A proof of the momentum center of edge-layer mode component}\label{AppendixC}

\begin{figure*}[t]
	\begin{center}
		\includegraphics[width=\linewidth]{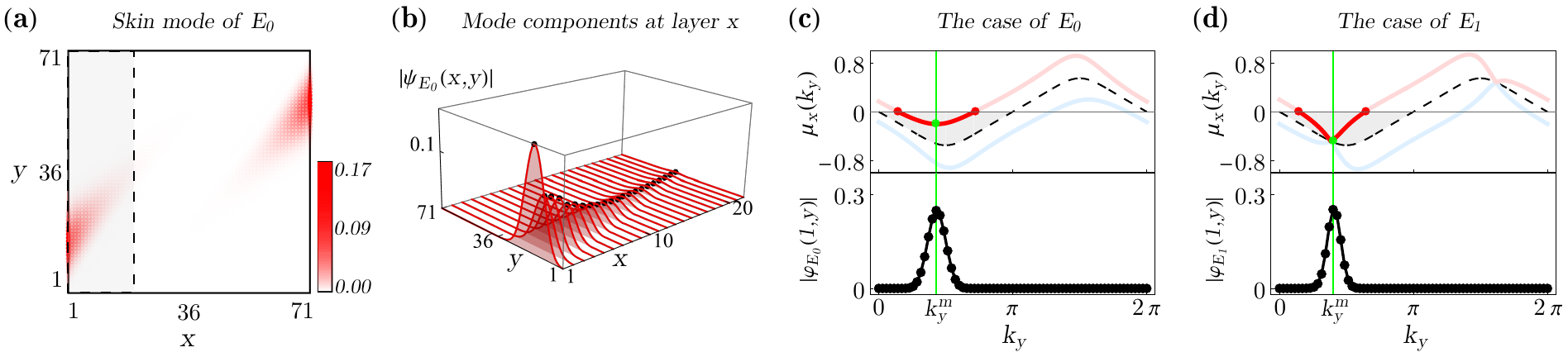}
		\par\end{center}
    \protect\caption{\label{fig:M1}~(a) shows the spatial profile of the skin mode with energy $E_0=0.877-0.751i$ under the square geometry of size $L_x=L_y=71$. 
    The Hamiltonian is given by Eq.(\ref{MT_Model}).  
    We extract the mode components in the first 20 $x$-layers (indicated by the gray region in (a)), and plot their amplitudes in (b). 
    It is shown in (b) that the mode components decay and their centers shift (as represented by the black dots) when moving away from the edge layer $x=1$. 
    In (c), the upper panel shows the corresponding solutions $\mu_{x,1}(E_0,k_y)$ (the blue curve), $\mu_{x,2}(E_0,k_y)$ (the red curve), and the cylinder-geometry GBZ $\tilde{\mu}_x(k_y)$ (the black dashed curve) for the non-Bloch Hamiltonian given by Eq.(\ref{MT_ModelCyl}). 
    Here, the green dot indicates the minimum of the red opaque arc, which terminates at two Fermi points indicated by the red dots. 
    The lower panel shows the numerical Fourier transform for the edge-layer component of the skin mode at $x=1$. 
    (d) follows the same procedure as that in (c), but it is calculated for the energy $E_1=-2.867-2.656i$. 
    The Hamiltonian parameters are set to $t=2,\gamma=1$.} 
\end{figure*}

Here, we offer a detailed proof that, for an edge-skin mode with energy $E_0$, the momentum center of its edge-layer component is fully determined by the non-Bloch bulk Hamiltonian $\mathcal{H}(\beta_{x},k_{y})$. 

We use the example given by Eq.(\ref{MT_Model}). 
The spatial profile of the skin mode with energy $E_0$, denoted as $\psi_{E_0}(\bm{r})$, under the square geometry is illustrated in Fig.~\ref{fig:M1}(a). 
The system size is $L_x=L_y=71$. 
The layer-$x_0$ component of the skin mode is denoted by $\psi_{E_0}(x_0,y)$. 
Therefore, the edge-layer components are $\psi_{E_0}(1,y)$ and $\psi_{E_0}(L_x,y)$ for left- and right-side $y$-edges, respectively. 
Here, we only focus on the edge-layer component $\psi_{E_0}(1,y)$. 
We extract the first 20 $x$-layers components near the edge layer $x=1$, indicated by the gray region surrounded by the black dashed box in Fig.~\ref{fig:M1}(a), and plot them in Fig.~\ref{fig:M1}(b). 
These components decay and their centers (the black dots in Fig.~\ref{fig:M1}(b)) shifts when away from the localized edge. 
When we consider the layer $x$ as time $t$, namely $(x,y)\rightarrow (t,y)$, the near-edge components of skin mode corresponds to a 1D finite-time wavepacket dynamics in the $y$ direction. 
The wavepacket decays and propagates with a velocity. 
Using this mapping, the skewness can be understood as the wavepacket group velocity in the 1D dynamics. 
In the finite-time dynamics, the wavepacket has not yet encountered the boundaries in the $y$ direction, therefore, the velocity (or skewness) should be captured by the Hamiltonian with PBC in $y$~\cite{MaoLiang2021} and OBC in $x$ direction, effectively being in a cylinder geometry. 

For the Hamiltonian in Eq.(\ref{MT_Model}), the non-Bloch Hamiltonian for the real-value edge momenta $k_y$ can be written as $\mathcal{H}(\beta_x,k_y)$ in Eq.(\ref{MT_ModelCyl}). 
For each fixed $k_y$, $\mathcal{H}(\beta_x,k_y)$ reduces to 1D non-Bloch Hamiltonian. 
Using the 1D non-Bloch theorem~\cite{Yao2018,Murakami2019_PRL}, the $k_y$-dependent generalized Brillouin zone (GBZ) can be obtained as 
\begin{equation}\label{M_kyGBZbeta}
    |\beta_{x,1}(E,k_y)| = |\beta_{x,2}(E,k_y)|,
\end{equation}
where ${\beta_{x,i}(E,k_y)}$ represents $i$-th solution of ${\det{[\mathcal{H}(\beta_x,k_y)-E]}=0}$ and has been ordered according to their amplitudes ${\beta_{x,i}(E,k_y)| \leq |\beta_{x,i+1}(E,k_y)|}$. 
The $k_y$-dependent GBZ condition can be satisfied only when $E$ is in the spectrum of Hamiltonian with PBC in $y$ and OBC in $x$ direction, namely cylinder-geometry spectrum. 
Alternatively, we express the $k_y$-dependent GBZ condition in terms of $\mu_x\equiv \log{|\beta_x|}$ as 
\begin{equation}\label{M_kyGBZmu}
    \tilde{\mu}_{x}(k_y) \equiv \mu_{x,1}(E,k_y) = \mu_{x,2}(E,k_y),
\end{equation}
with $\mu_{x,i}(E,k_y)\leq \mu_{x,i+1}(E,k_y)$. 
The $k_y$-dependent GBZ $\tilde{\mu}_{x}(k_y)$ is plotted as the black dashed curve in Fig.~\ref{fig:M1}(c) and (d). 

For the layer-$x_0$ component $\psi_{E_0}(x_0,y)$, we can always take the numerical Fourier transform. 
The corresponding momentum distribution at layer-$x_0$ can be calculated as 
\begin{equation}
    \varphi_{E_0}(x_0,k_y) = \frac{1}{\sqrt{L_y}}\sum\nolimits_{y=1}^{L_y} e^{-ik_y y} \, \psi_{E_0}(x_0,y).
\end{equation}
For example, the edge-layer momentum distribution $|\varphi_{E_0}(1,k_y)|$ is shown in Fig.~\ref{fig:M1}(c). 
Next, we prove that the edge momentum center $k_y^m$ of $\varphi_{E_0}(1,k_y)$, as indicated in Fig.~\ref{fig:M1}(c), is fully dependent on the non-Bloch bulk Hamiltonian $\mathcal{H}(\beta_x, k_y)$. 

We begin with the mode component at layer $x_0$, of which the momentum distribution is $\varphi_{E_0}(x_0,k_y)$. 
It is assumed that the layer $x_0$ is away from the edge layer, for example $x_0=20$. 
As illustrated in Fig.~\ref{fig:M1}(b), the amplitude of layer component will be amplifying when goes from the layer $x_0$ to the edge layer $x=1$, i.e., in the $-x$ direction. 
For each $k_y$ component, the amplification rate in the $-x$ direction can be given by the propagator: 
\begin{equation}\label{M_FormalPropa}
    \mathcal{G}_{E_0}(-x,k_y) = \frac{1}{2 \pi i}\oint_{C_{\beta_x}(k_y)} \frac{d \beta_x}{\beta_x} \frac{\beta^{-x}}{E_0^+ - \mathcal{H}(\beta_x,k_y)}, 
\end{equation}
where $E_0$ is the energy of skin mode, and $C_{\beta}(k_y)$ represents the $k_y$-dependent GBZ, as defined in Eq.(\ref{M_kyGBZbeta}) or Eq.(\ref{M_kyGBZmu}). 
Here, we emphasize that the skin mode is defined under fully open boundary conditions. However, we can effectively calculate the skewness under PBC in the $y$ and OBC in the $x$ direction. 
The reason is that the skin mode has almost faded away by the time it reaches the open boundary in the $y$ direction, therefore, the influence of the boundary conditions in the $y$ direction is negligible. 
Consequently, we use the $k_y$-dependent GBZ as the integral contour of the propagator in Eq.(\ref{M_FormalPropa}). 

\begin{figure*}[t]
    \begin{center}
    \includegraphics[width=1\linewidth]{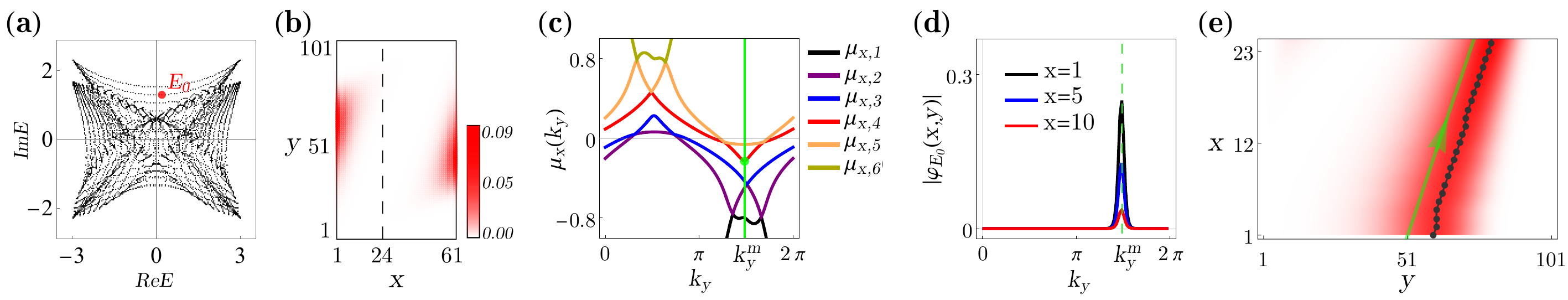}
    \par\end{center}
    \protect\caption{\label{fig:M2}~Skewness in a generic model given by Eq.(\ref{M_GeneModel}). 
    (a) represents the OBC eigenvalues of the Hamiltonian with system size $L_x=61$ and $L_y=101$. 
    (b) shows the spatial distribution of the skin mode with energy $E_0 = 0.212 + 1.295i$, as indicated by the red dot in (a). The black dashed line refers to the layer at $x = 24$. 
    (c) The solutions, $\mu_{x,i}(k_y)$, have been ordered according to their amplitudes. 
    The green intersection dot denotes the minimum of $\mu_{x,4}(k_y)$. 
    (d) The Fourier transform of the mode components at layers $x=1, 5, 10$ shows a constant momentum center. 
    (e) Correspondingly, the edge-skin mode with energy $E_0$ exhibits constant skewness, as denoted by the black dots, which is consistent with the analytical result $s_{\mathrm{edge}-y}(E_0) = 0.9784$, as indicated by the green arrow. } 
\end{figure*}

The propagator can be calculated according to Cauchy's integral theorem. 
For a given energy $E_0$, the integral value is determined by the roots of $E_0+i0^+ - \mathcal{H}(\beta_x,k_y)=0$. 
These contributing roots satisfy the following conditions: 
\begin{itemize}
    \item[(i)] ${|\beta_x(E_0,k_y)|<1}$ due to the amplification of mode components in the $-x$ direction;
    \item[(ii)] $\beta_x(E_0,k_y)$ is outside the $k_y$-dependent GBZ $C_{\beta}(k_y)$ according to the integral contour in Eq.(\ref{M_FormalPropa}). 
\end{itemize}
Therefore, the integral value of $\mathcal{G}_{E_0}(-x,k_y)$ is contributed by the roots that are inside the BZ ($|\beta_x(k_y)|\equiv 1$) and outside the $k_y$-dependent GBZ ($C_{\beta}(k_y)$). 
In terms of $\mu_x(E_0,k_y) = \log|\beta_x(E_0,k_y)|$, these contributing roots can be reformulated as: 
\begin{equation}
    \mathrm{M}_x(E_0,k_y):=\{\mu_x(E_0,k_y)|\tilde{\mu}_x(k_y) < \mu_x(E_0,k_y) < 0\}.
\end{equation}
As shown in Fig.~\ref{fig:M1}(c)(d), the region between BZ and $k_y$-dependent GBZ is represented by the gray region. 
The roots, $\mathrm{M}_x(E_0,k_y)$, contributing to the propagator $\mathcal{G}_{E_0}(-x,k_y)$ are indicated by the red opaque arc in this gray region. 
Note that the red opaque arc terminates at two Fermi points, as denoted by the red dots in Fig.~\ref{fig:M1}(c). 

In the large-size limit, each $k_y$ component in $\varphi_{E_0}(x_0,k_y)$ will amplify in the $-x$ direction. 
The $k_y$ component that gains the maximal amplification is contributed by the minimum in $\mathrm{M}_x(E_0,k_y)$, which is indicated by the green dot in Fig.~\ref{fig:M1}(c), expressed as:
\begin{equation}
    |\mathcal{G}_{E_0}(-x,k_y)| \propto e^{-\mu_{x,2}(k_y^m)x}.
\end{equation}
Consequently, the momentum center at the edge layer component must be $k_y^m$, where $\mathrm{M}_x(E_0,k_y)$ reaches its minimum. 
Two examples are presented as ${E_0=0.877 - 0.751 i}$, where ${E_0 \notin \sigma_{\text{cyl}}}$ is marked in Fig.~\ref{fig:3}(b), and ${E_1=-2.867 - 2.656 i}$, where ${E_1 \in \sigma_{\text{cyl}}}$ is marked by a black dot in Fig.~\ref{fig:3}(b). 
These correspond to Fig.~\ref{fig:M1}(c) and Fig.~\ref{fig:M1}(d), respectively.
In Fig.~\ref{fig:M1}(c)(d), the solutions $\mu_x(k_y,E_{0,1})$ are presented in the upper panels, and the edge-layer momentum distributions by a numerical Fourier transform are shown in the lower panels. These numerical calculations perfectly agree with our analytical results in both examples. 
So far, we have proved that the momentum center of edge-layer component is fully determined by the non-Bloch Hamiltonian $\mathcal{H}(\beta_x,k_y)$, without requiring any boundary details. 


\begin{figure*}[t]
	\begin{center}
		\includegraphics[width=\linewidth]{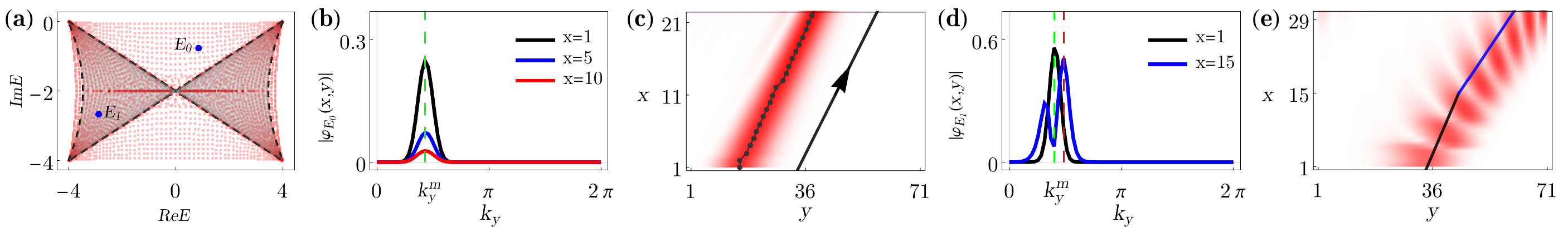}
		\par\end{center}
    \protect\caption{\label{fig:S3}~(a) shows the OBC eigenvalues (the red dots), including two specific energies $E_0=0.877-0.751i$, and $E_1=-2.867-2.656i$. 
    (b) presents the Fourier transform of mode components for the energy $E_0$ at layers $x=1, 5, 10$, demonstrating that the momentum center remains invariant as $x$ increases. 
    (c) Correspondingly, for the energy $E_0$, the edge-skin mode exhibits constant skewness, which is consistent with the analytical result indicated by the black arrow.
    (d) For the case of $E_1$, the Fourier transform of mode components at layer $x=1,15$ shows a shifting momentum center as $x$ increases. 
    Consequently, the skewness will bend when away from the edge $y$, as illustrated in (e). 
    } 
\end{figure*}

\section{The formula of skewness in generic cases}\label{AppendixD}

The formula of skewness can be straightforwardly extended into more generic cases with long-range hoppings. 
Without loss of generality, we assume that the skin mode under fully OBCs is localized on the edge $y$. 
To solve the skewness, the corresponding non-Bloch Hamiltonian with real-valued edge momenta can be expressed as 
\begin{equation}
    \mathcal{H}(\beta_x,k_y) = \sum_{l=-m}^n c_l(k_y) \beta_x^l,
\end{equation}
where $m$ and $n$ are finite integers representing the maximal rightward and leftward hopping range, respectively, and $c_l(k_y)$ indicates the $k_y$-dependent coefficient. 
For a given OBC eigenvalue $E_0$, we have $(m{+}n)$ solutions of $\mu_{x,i}(k_y)$, which can be ordered according to their magnitudes $\mu_{x,i}(k_y)\leq \mu_{x,i+1}(k_y)$. 
Therefore, the $k_y$-dependent GBZ can be expressed as 
\begin{equation}\label{M_GeneGBZ}
    \tilde{\mu}_x(k_y) \equiv \mu_{x,m}(E,k_y) = \mu_{x,m+1}(E,k_y).
\end{equation}
The condition is satisfied if and only if $E$ is in the corresponding cylinder-geometry spectrum ($\sigma_{\text{cyl}}$). It is the direct extension of Eq.(\ref{M_kyGBZmu}). 
Here, the cylinder geometry refers to the PBC in the $y$ and OBC in the $x$ direction. 
Therefore, for the case of $E_0 \notin \sigma_{\mrm{cyl}}$, the corresponding solutions can be ordered as:
\begin{equation}
    \begin{split}
    \mu_{x,1}(k_y) \leq \dots \leq \tilde{\mu}_x(k_y) < \mu_{x,m+1}(k_y) \leq \dots \mu_{x,m+n}(k_y).
    \end{split}
\end{equation}
It's important to note that when $E_0\notin \sigma_{\mrm{cyl}}$, for all $k_y$, $\mu_{x,m+1}(E_0,k_y) \neq \tilde{\mu}_x(k_y)$. 
In a similar manner, we can extend the conclusion regarding the edge-layer momentum center of the edge-skin mode to generic cases as follows: The momentum center of the edge-layer component is determined by the minimum of $\mu_{x,m+1}(E_0,k_y)$, denoted as $k_y^m$. 
Consequently, the edge-layer component satisfies
\begin{equation}
    \partial_{k_y}\mu_{x,m+1}(k_y^m) = 0
\end{equation}
for $E_0\notin \sigma_{\mrm{cyl}}$, which represents the generalization of Eq.(\ref{MT_StableCondition}). 
Accordingly, the skewness of edge-skin mode with energy $E_0$ can be further extended as 
\begin{equation}
    s_{\mrm{edge{-}y}}(E_0) = \partial_{k_y}k_{x,m+1}(k_y^m),
\end{equation}
where $k_{x,i}$ corresponds to $\mu_{x,i}$. 

Here, we present a generic numerical example to support the above generalization for the skewness of edge-skin mode. 
The non-Bloch Hamiltonian with real-valued edge momenta is given by 
\begin{equation}\label{M_GeneModel}
    \begin{split}
    \mathcal{H}(\beta_x,k_y) &= \frac{-i(\beta_x^3+\beta_x^{-3})}{3} + \frac{i (\beta_x+\beta_x^{-1})}{6} \\ 
    &+ (i-\frac{3}{2}e^{ik_y}) \beta_x^2 + (i-\frac{3}{2}e^{-ik_y})\beta_x^{-2}.
    \end{split}
\end{equation}
The bulk Hamiltonian respects reciprocity, $\mathcal{H}^T(\beta_x,k_y) = \mathcal{H}(1/\beta_x,-k_y)$. 
The OBC eigenvalues, calculated for a system size of $L_x=61$ and $L_y=101$, are represented by black dots in Fig.~\ref{fig:M2}(a). 
For illustration, we calculate the skewness of the skin mode with energy ${E_0 = 0.212 + 1.295 i}$ (the red dot in Fig.~\ref{fig:M2}(a)). 
The spatial distribution of the corresponding wavefunction is plotted in Fig.~\ref{fig:M2}(b), where the black dashed line indicates the layer at $x=24$. 
The wavefunction is symmetrically localized at the two edges in the $y$ direction. 
For the given $E_0$, we obtain the solutions $\mu_{x,i=1,\dots,6}(k_y)$ by solving the characteristic equation ${\mathcal{H}(\beta_x,k_y)-E_0=0}$. 
These solutions are ordered by their amplitudes, with $\mu_{x,i}(k_y)\leq \mu_{x,i+1}(k_y)$, and are illustrated in Fig.~\ref{fig:M2}(c) by the curves in different colors. 
Since $E_0\notin \sigma_{\mrm{cyl}}$, there exists a `gap' between $\mu_{x,3}(k_y)$ (the blue curve) and $\mu_{x,4}(k_y)$ (the red curve). 
The $k_{y}$-dependent GBZ, $\tilde{\mu}(k_y)\equiv \mu_{x,3}(E,k_y) = \mu_{x,4}(E,k_y) $, lies within this gap (not shown here). 

According to our conclusion, the momentum center of the edge-layer component is determined by the minimum of $\mu_{x,4}(k_y)$, as indicated by the green intersection dot in Fig.~\ref{fig:M2}(c). 
The corresponding momentum is denoted by $k_y^m$. 
At the momentum $k_y^m$, $\mu_{x,4}(k_y)$ reaches its minimum and satisfies the condition
\begin{equation}
    \partial_{k_y}\mu_{x,4}(k_y^m) = 0,
\end{equation}
which suggests that the edge momentum center remains unchanged as layer $x$ increases. 
This is verified in Fig.~\ref{fig:M2}(d), where the Fourier transform of layer components at $x=1, 5, 10$ is shown, and the momentum centers consistently lie at $k_y^m$. 
Therefore, the skewness of the skin mode with energy $E_0$ is formulated as
\begin{equation}
s_{\mathrm{edge{-}y}}(E_0) = \partial_{k_y}k_{x,4}(k_y^m),
\end{equation}
and analytically calculated to be ${s_{\mathrm{edge{-}y}}(E_0)=0.9784}$ for ${E_0 = 0.212 + 1.295 i}$. 
The analytical result, ${\delta y = s_{\mathrm{edge{-}y}}(E_0) \delta x}$, represented by the green arrow in Fig.~\ref{fig:M2}(e), aligns perfectly with the numerical calculations, which are denoted by the black dots in Fig.~\ref{fig:M2}(e). 

\section{The skewness within different energy spectrum regions}\label{AppendixE}

In this section, we show that the skewness of the edge-skin mode manifests differently for the OBC eigenvalues within and outside the cylinder-geometry spectrum region in the complex-energy plane. 

Here, we use the example given by Eq.(\ref{MT_Model}) as our demonstration. 
The open-boundary eigenvalues are evaluated and represented by the red dots in Fig.~\ref{fig:S3}(a). 
These OBC eigenvalues can be further divided into two sections
$$E_{\tx{OBC}}\notin \sigma_{\mathrm{cyl}} \,\,\, \tx{and} \,\,\, E_{\tx{OBC}}\in \sigma_{\mathrm{cyl}}.$$ 
Here, $\sigma_{\text{cyl}}$ signals the cylinder-geometry spectrum. 
As indicated in Fig.~\ref{fig:S3}(a), we select $E_0\notin \sigma_{\mathrm{cyl}}$ and $E_1\in \sigma_{\mathrm{cyl}}$ as two representative OBC eigenvalues to show the skewness of their corresponding skin modes. 

\begin{figure*}[t]
    \begin{center}
        \includegraphics[width=.8\linewidth]{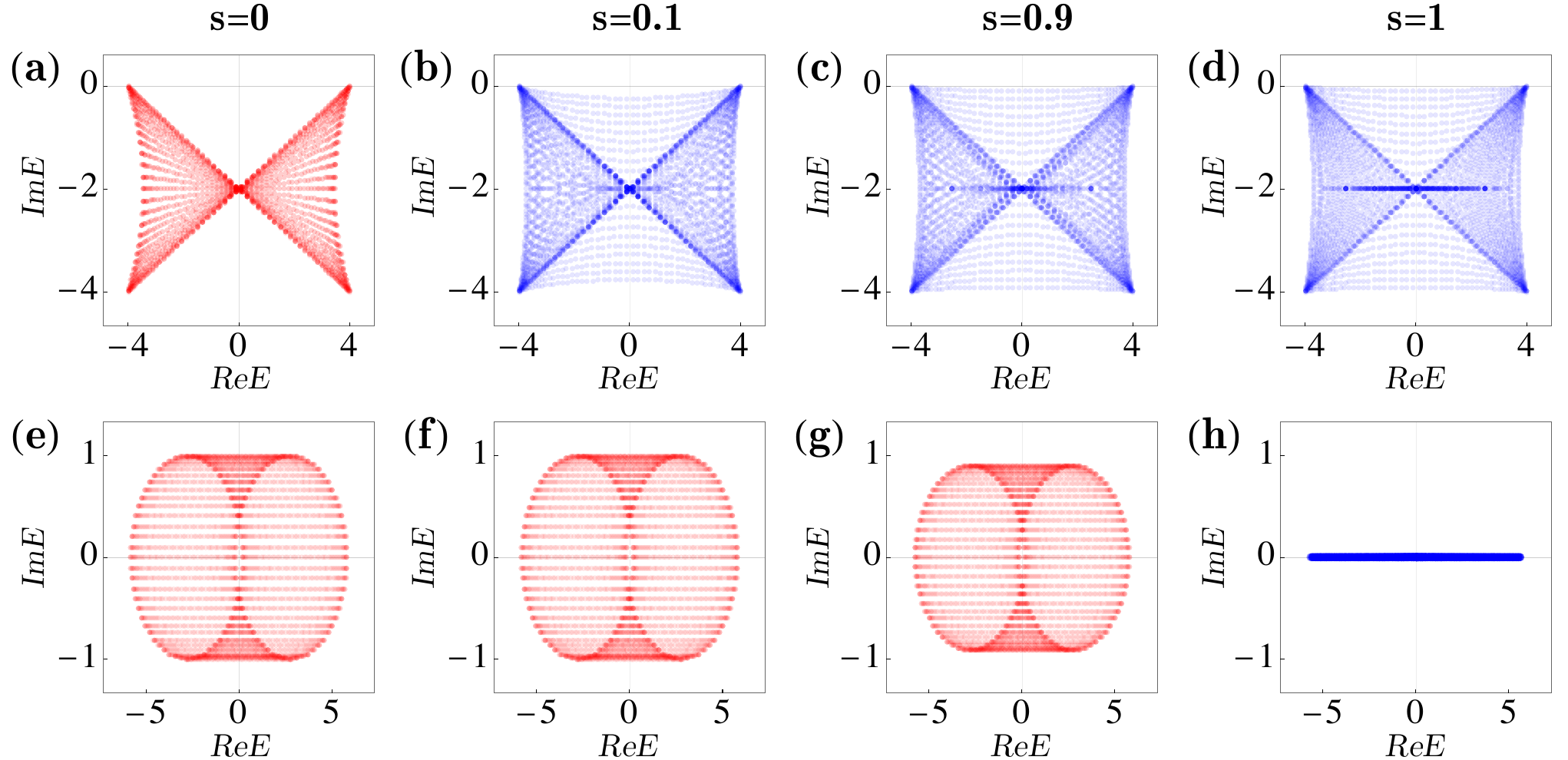}
        \par\end{center}
        \protect\caption{\label{fig:S4}~In the cases of the edge-skin effect (a)-(d) and corner-skin effect (e)-(h), the energy spectra change as the boundary conditions transition from cylinder geometry (a)(e) to fully open-boundary geometry (d)(h). 
        The non-perturbative phenomenon in the spectrum occurs near the cylinder geometry in the case of edge-skin effect, as demonstrated by (a) and (b), whereas it occurs near the fully open-boundary geometry in the case of corner-skin effect, as illustrated by (g) and (h).} 
\end{figure*}

In the case of $E_0\notin \sigma_{\mrm{cyl}}$, the solutions $\mu_{x,i}(k_y)$ can be obtained by solving ${\mathcal{H}(\beta_x,k_y)-E_0=0}$. These solutions have been ordered by $\mu_{x,i}(k_y)\leq \mu_{x,i+1}(k_y)$. 
At the edge momentum center $k_y^m$, it satisfies 
\begin{equation}
    \partial_{k_y}\mu_{x,2}(k_y^m) = 0.
\end{equation}
It means that wavepacket at the initial time (that is the edge-layer component) has a constant group velocity. 
In other words, the edge momentum center will not shift as $x$ increases. 
We show the Fourier transform for layer $x=1,5,10$ in Fig.~\ref{fig:S3}(b). 
It can be seen that the momentum center is invariant as layer $x$ increases. 
The constant group velocity, namely skewness, is calculated as 
\begin{equation}\label{SM_Skewness}
    s_{\tx{edge{-}y}}(E_0) = \partial_{k_y} k_{x,2}(k_y^m),
\end{equation}
As an example, we select $E_0=0.877 - 0.751 i$. 
The first 20 $x$-layer components is shown in Fig.~\ref{fig:S3}(c), where each layer component has been renormalized. 
The black dots represent the position of maximum amplitude for each layer component. 
According to formula of skewness in Eq.(\ref{SM_Skewness}), the skewness can be analytically calculated as $s_{\tx{edge{-}y}}(E_0) = 1.11279$. 
The analytical result, $\delta y-s_{\tx{edge{-}y}}(E_0) \delta x=0$, is indicated by the black arrow in Fig.~\ref{fig:S3}(c), which agrees well with the numerical results (the direction is given by the black dots). 

For the case of $E_1\in \sigma_{\mrm{cyl}}$, the solutions $\mu_{x,i}(k_y)$ always intersect with the $k_y$-dependent GBZ. 
The example is shown in Fig.~3(g) in the main text. 
This leads to 
\begin{equation}\label{SM_ZeroAcceE1}
    \partial_{k_y}\mu_{x,2}(k_y^m) \neq 0.
\end{equation}
What it means is that the edge momentum center will be shifted as layer $x$ increases. 
As a result, the declination of the skin mode varies as it shifts away from the edge layer. 
As an example, we select ${E_1=-2.867 - 2.656 i}$. 
As illustrated in Fig.~\ref{fig:S3}(d), the edge momentum center $k_y^m$ shifts into the other two momenta as layer $x$ changes from $x=1$ to $x=15$, which is indicated by the green dashed line for $x=1$ and red dashed line for $x=15$. 
For a clear comparison, the momentum distribution at different layers, namely $\varphi_{E_1}(1,k_y)$ and $\varphi_{E_1}(15,k_y)$, have been renormalized. 
Correspondingly, the skewness of the skin mode will bend when it moves away from the edge layer, as illustrated by the black and blue lines in Fig.~\ref{fig:S3}(e). 

\section{Anomalous spectral sensitivity in the edge-skin effect}\label{AppendixF}

In this section, we show that in the edge-skin effect, the spectrum under a specified cylinder geometry is highly sensitive to the perturbations that break the fragile reciprocity, such as the weak onsite impurities. 
For example, when the edge $y$ is the localized edge of the skin modes, we specify the cylinder geometry as that with PBC in the $y$ and OBC in the $x$ direction. 
Based on this, the reciprocity symmetry in the edge-skin effect, given by Eq.(\ref{SM_SymmEdgeSkin}), further requires the characteristic equation $f_E(k_x,\mu_x,k_y)=\det{[\mathcal{H}(\beta_x=e^{ik_x+\mu_x},k_y)-E]}$ to satisfy 
\begin{equation}
    f_E(k_x,\mu_x,k_y) = f_E(-k_x,-\mu_x,-k_y) = 0,
\end{equation}
which is termed fragile reciprocity. 
The fragile reciprocity can be easily broken by the perturbations that break the translation symmetry in the $y$ direction, such as on-site disorders or open boundaries, thus termed `fragile'. 

Here, we numerically demonstrate that the spectral sensitivity is anomalous and unique to the edge-skin effect, distinguishing it from the corner-skin effect. 
The model Hamiltonian of the edge-skin effect is given by
\begin{equation*}
h_{\mathrm{ESE}}(\bm{k}) = 4\cos(k_x+k_y) - 2 i (1-\cos{k_x}).
\end{equation*}
We start with the Hamiltonian under the cylinder geometry, that is, PBC in $y$ and OBC in $x$ direction. 
Modulating the strength of the boundary link $s$ in the $y$ direction allows for a transition from cylinder geometry ($s=0$) to fully OBC geometry ($s=1$). 
Therefore, the set of Hamiltonians is formally expressed as 
\begin{equation}
    H_s = H_{\mrm{cyl}} - s H_{\mrm{B}},
\end{equation}
where $H_{\mathrm{cyl}}$ denotes the cylinder-geometry Hamiltonian, and $H_{\mathrm{B}}$ represents the boundary link term. 
Therefore, $H_0$ refers to the Hamiltonian under cylinder geometry, and $H_1$ corresponds to the Hamiltonian under fully OBCs. 
The spectra corresponding to different boundary links are depicted in Fig.~\ref{fig:S4}(a)-(d). 
Remarkably, a slight modification in the boundary link (${s=0.1}$) can result in a substantial changes to the spectrum, as shown by the comparison of Fig.~\ref{fig:S4}(a) and (b).  
After this non-perturbative change of the spectrum, the energy spectrum in Fig.\ref{fig:S4}(b) is very close to the fully open-boundary spectrum shown in Fig.\ref{fig:S4}(d). 

The Hamiltonian for the corner-skin effect is given by
\begin{equation*}
h_{\mathrm{CSE}}(\bm{k}) = 3(\cos{k_x}+\cos{k_y}) + i (\sin{k_x}+\sin{k_y}).
\end{equation*}
The energy spectra corresponding to different boundary links are illustrated in Fig.~\ref{fig:S4}(e)-(h). 
Compared with the edge-skin effect in Fig.~\ref{fig:S4}(a)-(d), in the corner-skin effect, the minor modification of boundary link can be considered as a perturbation to its energy spectra, as displayed in Fig.~\ref{fig:S4}(e)-(g). 
The non-perturbative effect in the spectrum occurs near the fully OBC spectrum~\cite{ZSaGBZPRL,LiLH2020_NC}, as illustrated in Fig.~\ref{fig:S4}(g) and (h).  

\bibliography{refs}
\bibliographystyle{apsrev4-2}
\end{document}